\documentclass[prb,twocolumn,amsmath,amssymb,superscriptaddress,citeautoscript,longbibliography]{revtex4-2}

\usepackage{graphicx,color}
\usepackage{lipsum}
\usepackage{braket}
\usepackage[colorlinks,urlcolor=blue,citecolor=blue]{hyperref}
\usepackage[utf8]{inputenc}

\begin{document}

\title{Systematic compactification of the two-channel Kondo model. \\II.
\textit{Comparative study of scaling and universality}}

\author{Aleksandar \surname{Ljepoja}}
\affiliation{Department of Physics, University of Cincinnati, OH-45221, USA}
\author{Nayana \surname{Shah}}
\affiliation{Department of Physics, Washington University in St.~Louis, MO-63160, USA}
\author{C.~J.~\surname{Bolech}}
\affiliation{Department of Physics, University of Cincinnati, OH-45221, USA}
\affiliation{ITAMP, Harvard-Smithsonian Center for Astrophysics, Cambridge, Massachusetts 02138, USA}

\begin{abstract}

Following up on the systematic compactification of the two-channel Kondo model (and its multichannel extensions; see \href{https://doi.org/10.48550/arXiv.2308.03569}{companion paper I}) and the demonstration of its validity over the past proposal of compactification, we resort to a study of scaling using Anderson's simple poor man's procedure to carry out a comparative study of these two and the original model. By doing so we unveil a \textit{universal} agreement among the three models in how they flow upon scaling, and suggest the general limits of such a concordance. In this way we further elucidate the conditions under which the standard simplifications implicit in many bosonization-based mappings (particularly of quantum impurity models) can be used reliably, and when the consistent bosonization-debosonization approach is needed.
\end{abstract}

\maketitle

\section{Introduction}

A variety of perturbative and nonperturbative methods have been developed and employed to access the evasive intermediate-coupling fixed point of the two-channel Kondo model. One of them, the \textit{compactification} of Kondo-type impurity models, has relied on spin-charge separation and the fact that the interaction involves only the spin sector (and not the isospin; see below) in an attempt to shift the fixed point to strong coupling. 
The original proposal along these lines was nonconstructive and based entirely on symmetry considerations \cite{coleman1995b}. It was a pedestrian approach to the compactification of quantum impurity models that starts from the observation that the full local symmetry of the bulk degrees of freedom  (assuming particle-hole symmetry at half-filling and no gauge fields) for a single species of spin-half fermions is $SO(4)\sim SU\!(2)^\text{spin}\times SU\!(2)^\text{iso}$ \cite{coleman1995b,Zhang1996,*Zhang1996a}. Besides the spin algebra in the singly occupied sector, one has an additional iso-spin algebra between the empty and doubly occupied states of the local Hilbert space ---also sometimes referred as the $\eta$-spin in the context of the Hubbard model \cite{Yang1989}. (\textit{Nota bene}~that this high degree of symmetry can be made more manifest and better exploited using a Majorana description of the bulk fermions \cite{maldacena1997,Zhang1999}.) These two algebras are then coupled to the impurity \textit{in lieu} of the spin algebras of two independent channels. The claim is that the impurity-spin dynamics is preserved despite having now a single spin-full channel and, in that sense, the resulting model being more \textit{compact}.

A complementary way to define compactification could be to compare two impurity models with the same number of degrees of freedom in the bulk and say that an impurity-model interaction is more ``compact'' than another if a larger fraction of the symmetry generators of the bulk are involved in the interaction with the impurity. In the above discussion of compactifiction, the number of bulk degrees of freedom changed: one carried out an \textit{ad hoc} construction of a \textit{compactified} two-channel Kondo model by starting from the usual model and substituting the electron spin densities of \textit{both} channels by the spin and isospin densities of a \textit{single} channel 
\footnote{A different construction starts from the ph-symmetric single-channel Anderson model, rewritten in terms of Majorana fermions, for then arbitrarily changing the hybridization term to break the $SO(4)$ symmetry down to $SO(3)$ \cite{Zhang1996,*Zhang1996a}. This gives a model that shares some features of the two-channel Kondo model, but does not start from a \textit{bona fide} two-channel Anderson model \cite{bolech2006a,*iucci2008}. We shall leave aside those connections to be explored elsewhere.}.
As a result, the number of generators involved in the interaction with the impurity remains the same, but the fraction increases because the total number of generators is now half as large.

In the compactified model, two physically distinct $SU\!(2)$ densities 
\footnote{Connected to two separate level-1 \textit{current algebras} that can, however, be mapped to each other via a staggered particle-hole transformation that mixes sectors with different boundary conditions \cite{Zhang1999}.}
of a single channel compete to screen the impurity. Assuming their independence at the level of the noninteracting theory, (cf.~the standard Sugawara construction \cite{affleck1990,affleck1991,Affleck1992}), would guarantee that their correlators emulate the joint spin-density correlations of the original two-channel Kondo model \cite{coleman1995b}. However, unlike in the original case, the spin and isospin densities cannot both be \textit{active} simultaneously at the ``local site'' of an impurity \cite{coleman1995}, since the band spin and isospin are nonzero in separate sectors of the local Hilbert space. These two observations are in tension, highlighting the limits of the construction. Since in a compactified model the impurity would locally couple different symmetry-representation sectors of a single channel of bulk fermions, the properties of the bulk electrons would be affected differently as compared with the noncompact model
\footnote{In the language of \textit{boundary-condition-changing operators}, the impurity behaves differently in connection with the ``two-channel'' or ``compact'' bulks, respectively; cf.~Ref.~\onlinecite{affleck1994,*shah2003}.}.
On the other hand, the impurity dynamics is expected to stay the same, but the spin/isospin alternation was argued to shift the non-Fermi-liquid low-temperature fixed point towards infinite coupling and render channel-asymmetry from a relevant into a marginal perturbation \cite{coleman1995}.

The original and compact versions of the two-channel Kondo model were thus expected to share most of the intriguing aspects of the low-temperature impurity physics while differing at the details in welcomed ways 
\footnote{It was hoped that compactification would provide a blue print for stabilizing the non-Fermi-liquid physics against the otherwise-relevant channel asymmetry. This prospect found a match in the more recent proposals of topological Kondo impurities \cite{Beri2012,Altland2013,Altland2014,Li2023,*Koenig2023} (of high theoretical interest but facing steep experimental challenges) or of Kondo physics in correlated bulks \cite{Fiete2008,Koenig2020}, and in connection with more general $\mathfrak{o}(N)$-Kondo models.}.
Moreover, a number of subsequent studies [including the $\mathfrak{o}(3)$-Anderson] gave partial support to these ideas \cite{bulla1997,bulla1997a,ye1998}, but the ultimate fate of the fixed point stability remained largely unsettled. Interestingly and around the same time, the compactification of the two-channel Kondo model was recast as a precise correspondence between the original and compact versions of the model via a bosonization-debosonization (BdB)-based mapping \cite{schofield1997}. But since BdB can be exact under certain conditions (which are the same as those assumed for compactification), the original and compact models could be expected to be identical in all aspects of the impurity physics. The fact that they are not, poses a conundrum that we addressed in the first part of this study \cite{Ljepoja2024a}. We resolved it by pointing out the limitations of the \textit{conventional} approach to BdB (and thus to compactification), and we then used \textit{consistent} BdB \cite{shah2016,*bolech2016} to give a more nuanced compact model, which we argued to indeed be exactly equivalent to the original one and to go into the conventionally compactified one only after a mean-field-type reduction. We supported this by studying a number of (nongeneric) special limits in which the models can be handled without any approximations. We did not, however, address the full models in their more physically interesting regimes and the physics of their fixed points. The purpose of the present article is exactly that.

In what follows, we shall study the Hamiltonians of the original multichannel Kondo model and its different compactifications with a focus on their scaling properties. On physical grounds, the underlying expectation is that the scaling of the Kondo coupling

is a manifestation of observable physics that should stay (relatively) unchanged under model reparametrizations; in particular, it should not change at all if the BdB-based mapping is indeed an exact correspondence. In fact, the study of the crossover between UV and IR fixed points is what came to be known as the \textit{Kondo problem} and played a pivotal role in the development of renormalization group (RG) ideas \cite{wilson1975}. The rest of this work is thus organized as follows. (i) In the next section we present the standard picture of scaling. We first go over the technical aspects using the original formulation of the model and doing it with more detail than usual in order to be able to highlight the important differences later. Then we repeat the calculations for the compact versions of the model (taking the agnostic point of view that they are different \textit{schemes} for studying the same physics, despite the different Hamiltonians). (ii) In the following section we compare the results from the different schemes and discuss the universal and nonuniversal aspects of the physics. (iii) Finally, in the last section, we put things into further perspective, comment on what questions still remain open, and conclude.

\section{Poor man's scaling}

Along the way to solving the Kondo problem (finally achieved via a numerical implementation of the full RG program \cite{wilson1975} followed by a \textit{tour-de-force} Bethe ansatz solution \cite{andrei1983}), a rich set of innovative ideas and approaches were developed for the study of strongly interacting quantum systems. Among those, P.\,W.~Anderson pioneered the use of scaling ideas in a series of papers that culminated with his introduction of the poor man's approach to scaling \cite{Anderson1970}. This insightful work anticipated the development of the modern Wegner-Wilson momentum-shell RG \cite{Wegner1972,*Wegner1973,Wilson1974}. Here we give the details of the poor man's scaling calculation for the (multichannel) Kondo model in all the three versions of it discussed previously (namely, the direct formulation and the two possible compactified ones) \cite{Ljepoja2024a}, and up to third order in perturbation theory.

\subsection{Direct model formulation}
Our starting point is thus the (multichannel) Kondo Hamiltonian written in terms of the adiabatic Landau quasi-particles describing a Fermi-liquid metallic host interacting with a single spin-1/2 magnetic impurity. It is given by the sum of the \textit{free} conduction electron Hamiltonian ($H_0$) and the Kondo interaction terms
\begin{equation*}
\begin{split}
H_{K} = \sum_{\alpha,\sigma}\sum_{\vec{k}_{1}, \vec{k}_{2}} \bigg ( J_{\sigma}S^{\sigma}c^{\dagger}_{\vec{k}_{1} \bar{\sigma}\alpha}c_{\vec{k}_{2} \sigma \alpha} + \sigma J_{z}S_{z} c^{\dagger}_{\vec{k}_{1} \sigma \alpha}c_{\vec{k}_{2} \sigma \alpha} \bigg )
\end{split}
\end{equation*}
To carry out the \textit{scaling}, we are going to eliminate high-energy excitations in the conduction band (edge states), and find out how do the parameters of the Hamiltonian ($J_{z}$ and $J_{\perp}=J_{+}, J_{-}$) change under such eliminations \cite{Anderson1970,Hewson,Nevidomskyy2015}. Explicitly, we are going to project the eigenfunction of the Hamiltonian into three parts: $\psi_{0}, \psi_{1}$ and $\psi_{2}$; so that the eigenvalue equation becomes
\begin{equation}
\begin{pmatrix}
H_{00} & H_{01} & H_{02}\\
H_{10} & H_{11} & H_{12}\\
H_{20} & H_{21} & H_{22}
\end{pmatrix}
\begin{pmatrix}
\psi_{0}\\
\psi_{1}\\
\psi_{2}
\end{pmatrix}
= E
\begin{pmatrix}
\psi_{0}\\
\psi_{1}\\
\psi_{2}
\end{pmatrix}
\end{equation}
Here $\psi_{1}$ is a wave function component that describes a state in which there are no electrons in the upper band edge and no holes in the lower band edge (\textit{i.e.}, no high-energy excitations); $\psi_{0}$ corresponds to a state in which we have at least one hole in the lower band edge; and $\psi_{2}$ is the component of the wave function where there is at least one electron in the upper band edge. The Fermi energy is set to be zero so that the single electron states with energy in the range of $-D<\epsilon<0$ are occupied and states with energies $0<\epsilon<D$ are unoccupied (at zero temperature and in the absence of the impurity for a conduction band of bandwidth $2D$). Band-edge states are states with energies $-D<\epsilon<-D+|\delta D|$ or $D-|\delta D|<\epsilon< D$. The elements of the Hamiltonian matrix are organized so that $H_{nn^{\prime}}$ are the terms of the Hamiltonian that connect states $\psi_{n}$ and $\psi_{n^{\prime}}$, where $n,n^{\prime} = 0,1,2$. We are going to neglect the $H_{02}$ and $H_{20}$ elements of the effective Hamiltonian, because they do not contribute to the leading order in ${1}/{D}$. Eliminating $\psi_{0}$ and $\psi_{2}$ from the eigenvalue equation we get the effective eigenvalue equation for $\psi_{1}$:
\small
\begin{equation}\label{eq:PMscalingH}
\bigg [H_{11}+H_{12}(E-H_{22})^{-1}H_{21} + H_{10}(E-H_{00})^{-1}H_{01} \bigg ]\psi_{1} = E\psi_{1}
\end{equation}
\normalsize
where, in particular, $H_{21}$ labels the component of the Hamiltonian which creates a particle in an upper-band-edge single-particle state with momentum $\vec{q}$. It is given by
\begin{equation*}\label{particleH}
\begin{split}
H_{21} = & \sum_{\alpha}\sum_{\sigma}  \sum_{\vec{k}, \vec{q}} \bigg ( J_{\sigma}S^{\sigma}c^{\dagger}_{\vec{q} \bar{\sigma} \alpha}c_{\vec{k} \sigma \alpha}+  \sigma J_{z} S^{z} c^{\dagger}_{\vec{q} \sigma \alpha}c_{\vec{k} \sigma \alpha} \bigg )
\end{split}
\end{equation*}
On the other hand, the part of the Hamiltonian that creates a hole in a lower-band-edge single-particle state is given by $H_{01}$,
\begin{equation*}
\begin{split}
H_{01} = &  \sum_{\alpha}\sum_{\sigma} \sum_{\vec{k}, \vec{q}} \bigg ( J_{\sigma}S^{\sigma}c^{\dagger}_{\vec{k} \bar{\sigma} \alpha}c_{\vec{q} \sigma \alpha} + \sigma J_{z}S^{z} c^{\dagger}_{\vec{k} \sigma \alpha}c_{\vec{q} \sigma \alpha} \bigg )
\end{split}
\end{equation*}
and again we used $\vec{q}$ to label the edge state. The reverse processes are given by the Hermitian conjugates, $H_{12}$ and $H_{10}$, respectively. 

Now we can integrate out those edge states and observe what is happening with the coupling constant. This is the essence of the scaling procedure.

\subsubsection{Second-order poor man's scaling}

\begin{figure}[h]
\includegraphics[width=0.47\textwidth]{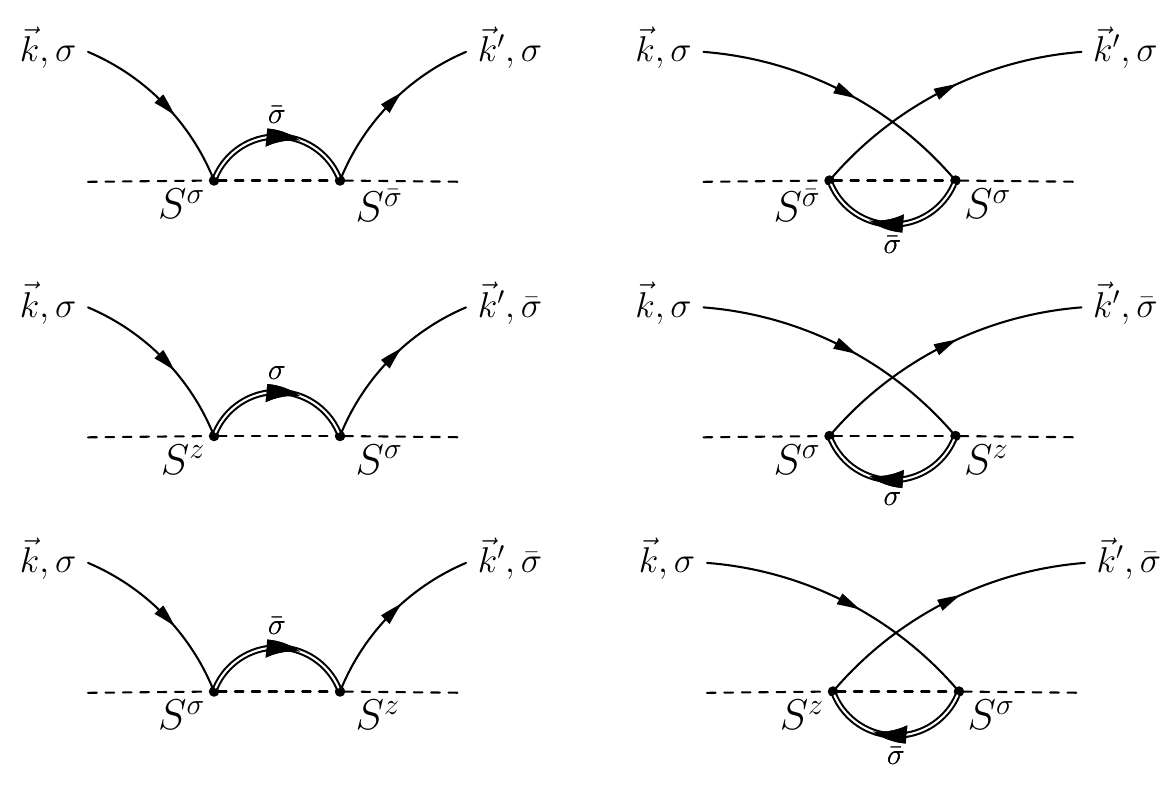}
\caption{Second-order processes for poor man's scaling. The dashed line represents the impurity, full lines are midband fermions, and the double line is a scattered edge-state fermion. The two diagrams in the first row contribute to the scaling of the $J_{z}$ coupling constant. The four other diagrams contribute to the scaling of $J_{\perp}$.}
\label{fig:2ndPMD}
\end{figure}

The so called T-matrix diagrams, depicting the different contributions to scaling, are given in Fig.~\ref{fig:2ndPMD} to second order in the coupling constants. All the diagrams in the left column are what we call ``particle'' diagrams. Those are the processes that are generated by the second term in the effective Hamiltonian of Eq.~(\ref{eq:PMscalingH}). They create and then destroy destroy a particle (\textit{i.e.}, an electron) in the upper band edge. Similarly, the diagrams on the right column correspond to the so called ``hole'' processes, given by the third term in the effective Hamiltonian, and they create and destroy a hole in the lower band edge.

For example, the contribution of the topmost left diagram in Fig.~\ref{fig:2ndPMD} is explicitly given by
\begin{equation*}
J^{2}_{\perp}\sum_{\alpha, \alpha^{\prime}}\sum_{\sigma, \sigma^{\prime}}\sum_{\vec{q},\vec{q}^{\prime}} \sum_{\vec{k},\vec{k}^{\prime}} S^{\bar{\sigma}}S^{\sigma}c^{\dagger}_{\vec{k}^{\prime} \sigma^{\prime} \alpha^{\prime}}c^{\ }_{\vec{q}^{\prime} \bar{\sigma}^{\prime} \alpha^{\prime}}\frac{1}{(E-H_{22})}c^{\dagger}_{\vec{q} \bar{\sigma} \alpha}c^{}_{\vec{k} \sigma \alpha}
\end{equation*}
where $\vec{q}$ labels the momentum of a particle in the band edge. We can approximate  $H_{22}$  with the noninteracting conduction electron Hamiltonian, and in that way we  keep only the second-order processes,
\begin{equation}
H_{22}\approx H_{0} = \sum_{\alpha}\sum_{\sigma}\sum_{\vec{k}}\epsilon_{\vec{k}}c^{\dagger}_{\vec{k} \sigma \alpha}c^{}_{\vec{k} \sigma \alpha}
\end{equation}
Assuming that the top band edge is unoccupied in the initial and final states, we can replace 
$c_{\vec{q}^{\prime} \bar{\sigma}^{\prime} \alpha^{\prime}}c^{\dagger}_{\vec{q} \, \bar{\sigma} \alpha} \to \delta_{\vec{q}^{\prime}\vec{q}} \, \delta_{\bar{\sigma}^{\prime} \bar{\sigma}} \, \delta_{\alpha^{\prime} \alpha}$. In that case the expression becomes
\begin{equation*}
J^{2}_{\perp} \sum_{\alpha} \sum_{\sigma} \sum_{\vec{q}} \sum_{\vec{k},\vec{k}^{\prime}} S^{\bar{\sigma}}S^{\sigma}c^{\dagger}_{\vec{k}^{\prime} \sigma \alpha}c^{\ }_{\vec{k} \sigma \alpha}\frac{1}{(E-\epsilon_{\vec{q}}+\epsilon_{\vec{k}}-H_{0})}
\end{equation*}
If the energy $E$ is measured relative to the ground state of the conduction electrons, then one sets $H_{0}=0$. Doing a summation over $\vec{q}$ and approximating $\epsilon_{q} \equiv D$, --which is to say that the band-edge width, $|\delta D|$, is infinitesimally small--, we get
\begin{equation}\label{eq:twoterms}
J^{2}_{\perp} \sum_{\alpha} \sum_{\sigma} \sum_{\vec{k},\vec{k}^{\prime}}\bigg (\frac{1}{2}- \sigma S_{z} \bigg ) c^{\dagger}_{\vec{k}^{\prime} \sigma \alpha}c_{\vec{k} \sigma \alpha}\frac{\rho_{0}|\delta D|}{(E-D+\epsilon_{\vec{k}})}
\end{equation}
where we have used the spin identity $S^{\bar{\sigma}}S^{\sigma}=\frac{1}{2}- \sigma S_{z}$, as well as the fact that $\sum_{q} = \rho_{0}|\delta D|$, where $\rho_{0}$ is the free-electron density of states [for the 1D case with $\epsilon_{k}=v_Fk$, the density of states is a constant, $\rho_{0}=1/(2\pi v_F)$; per spin, per channel]. There are thus two terms in the expression above. One is the potential-scattering term given by
\begin{equation}\label{potentialP}
\frac{1}{2}J^{2}_{\perp}\sum_{\alpha} \sum_{\sigma} \sum_{\vec{k}, \vec{k}^{\prime}}c^{\dagger}_{\vec{k}^{\prime} \sigma \alpha}c_{\vec{k} \sigma \alpha}\frac{\rho_{0}|\delta D|}{(E-D+\epsilon_{\vec{k}})}
\end{equation}
which will be important later when we discuss the so-called \textit{wave-function renormalization}. The second term in Eq.~(\ref{eq:twoterms}) will contribute to the rescaling of the $J_{z}$ coupling constant and is given by
\begin{equation}\label{eq:potentialP}
-J^{2}_{\perp} \sum_{\alpha} \sum_{\sigma} \sum_{\vec{k},\vec{k}^{\prime}} \sigma S_{z}c^{\dagger}_{\vec{k}^{\prime} \sigma \alpha}c_{\vec{k} \sigma \alpha}\frac{\rho_{0}|\delta D|}{(E-D+\epsilon_{\vec{k}})}
\end{equation}

\noindent This was the contribution of the ``particle'' process. Its ``hole'' counterpart is given by the topmost right diagram in Fig.~\ref{fig:2ndPMD}. As already said, we get this diagram from the third term in the effective Hamiltonian and it is explicitly given by
\begin{equation*}
J^{2}_{\perp} \sum_{\alpha, \alpha^{\prime}}\sum_{\sigma, \sigma^{\prime}} \sum_{\vec{q},\vec{q}^{\prime}} \sum_{\vec{k},\vec{k}^{\prime}} S^{\sigma}S^{\bar{\sigma}}c^{\dagger}_{\vec{q}^{\prime} \bar{\sigma}^{\prime} \alpha^{\prime}}c_{\vec{k}^{\prime} \sigma \alpha}\frac{1}{(E-H_{00})}c^{\dagger}_{\vec{k} \sigma \alpha}c_{\vec{q}^{}\bar{\sigma} \alpha}
\end{equation*}
Notice that now we have $H_{00}$ in the denominator. This is because this diagram creates, and destroys a hole in the lower band-edge with momentum $\vec{q}$. Again, assuming that $H_{00} \approx H_{0}$ and making the equivalent assumptions to those that we used for the particle diagram, we arrive at
\begin{equation}
J^{2}_{\perp}\sum_{\alpha}\sum_{\sigma}\sum_{\vec{k},\vec{k}^{\prime}}\bigg (\frac{1}{2}-\bar{\sigma} S_{z} \bigg ) c_{\vec{k}^{\prime} \sigma \alpha}c^{\dagger}_{\vec{k} \sigma \alpha}\frac{\rho_{0}|\delta D|}{(E-D-\epsilon_{\vec{k}})}
\end{equation}
Once again we have a potential scattering term,
\begin{equation}\label{eq:potentialH}
\frac{1}{2} J^{2}_{\perp}\sum_{\alpha} \sum_{\sigma} \sum_{\vec{k},\vec{k}^{\prime}}c_{\vec{k}^{\prime} \sigma \alpha}c^{\dagger}_{\vec{k} \sigma \alpha}\frac{\rho_{0}|\delta D|}{(E-D-\epsilon_{\vec{k}})}
\end{equation}
as well as a term that will rescale the parallel coupling constant,
\begin{equation}
-J^{2}_{\perp} \sum_{\alpha} \sum_{\sigma} \sum_{\vec{k},\vec{k}^{\prime}}\bar{\sigma}S_{z}c_{\vec{k}^{\prime} \sigma \alpha}c^{\dagger}_{\vec{k} \sigma \alpha}\frac{\rho_{0}|\delta D|}{(E-D-\epsilon_{\vec{k}})}
\end{equation}
To second order, the two potential-scattering contributions, Eqs.~(\ref{eq:potentialP}) and (\ref{eq:potentialH}),  combine into a constant energy shift; since the operators combine into a fermion anticommutator, which is a c-number. Thus, impurity potential scattering is indeed not generated, as expected from particle-hole symmetry. The c-number shift is going to lead to wave-function renormalization, as mentioned before, when we move on to the third-order calculation of scaling. 

For now, we focus on the terms that rescale the parallel Kondo interaction. Under the assumption that $\epsilon_{\vec{k}}, E << D$, and adding the two terms that rescale $J_{z}$ in a single expression, we arrive at
\begin{equation*}
\rho_{0} J^{2}_{\perp} \frac{|\delta D|}{D}\sum_{\alpha} \sum_{\sigma} \sum_{\vec{k},\vec{k}^{\prime}}S_{z}\bigg ( \sigma c^{\dagger}_{\vec{k} \sigma \alpha}c^{\ }_{\vec{k}^{\prime} \sigma \alpha} + \bar{\sigma} c^{\ }_{\vec{k}^{\prime} \sigma \alpha}c^{\dagger}_{\vec{k} \sigma \alpha} \bigg )
\end{equation*}
which, after commuting $c_{\vec{k}^{\prime},\sigma}$ and $c^{\dagger}_{\vec{k},\sigma}$, (the constant energy shift cancels between the two spin projections), will lead to the expression
\begin{equation}
2 \rho_{0} J^{2}_{\perp} \frac{|\delta D|}{D} \sum_{\alpha} \sum_{\sigma}\sum_{\vec{k},\vec{k}^{\prime}} \sigma S_{z}c^{\dagger}_{\vec{k} \sigma \alpha}c_{\vec{k}^{\prime} \sigma \alpha}
\end{equation}
and this implies the rescaling of the $J_{z}$ coupling constant:
\begin{equation}
J_{z}  \to J^{\prime}_{z} = J_{z}+2\rho_{0}\frac{|\delta D|}{D}J^{2}_{\perp}
\end{equation}

The last four diagrams in Fig.~\ref{fig:2ndPMD} are processes that rescale the $J_{\perp}$ coupling constant. The contraction and summation over the edge-state momentum proceed in the same way as in the diagrams we just discussed. However, we have a different spin structure at the vertices of the diagrams. Taking those two things into account we get that the contribution of the middle two diagrams is
\begin{equation*}
\begin{split}
& - \rho_{0} J_{\perp}J_{z}\frac{|\delta D|}{D} \sum_{\alpha} \sum_{\sigma} \sum_{\vec{k},\vec{k}^{\prime}}\sigma \bigg ( S^{\sigma}S^{z}- S^{z}S^{\sigma} \bigg ) c^{\dagger}_{\vec{k}^{\prime} \bar{\sigma} \alpha}c_{\vec{k} \sigma \alpha}\\
& \qquad = \rho_{0} J_{\perp}J_{z}\frac{|\delta D|}{D} \sum_{\alpha} \sum_{\sigma} \sum_{\vec{k},\vec{k}^{\prime}}\sigma^{2} S^{\sigma} c^{\dagger}_{\vec{k}^{\prime} \bar{\sigma} \alpha}c^{\ }_{\vec{k} \sigma \alpha}\\
& \qquad =\rho_{0} J_{\perp}J_{z}\frac{|\delta D|}{D}\sum_{\alpha} \sum_{\sigma} \sum_{\vec{k},\vec{k}^{\prime}} S^{\sigma} c^{\dagger}_{\vec{k}^{\prime} \bar{\sigma} \alpha}c^{\ }_{\vec{k} \sigma \alpha}
\end{split}
\end{equation*}
Moving to the last two diagrams, we obtain again the same final expression (one just arrives to it in a slightly different way),
\begin{equation*}
\begin{split}
& - \rho_{0} J_{\perp}J_{z}\frac{|\delta D|}{D} \sum_{\alpha} \sum_{\sigma} \sum_{\vec{k},\vec{k}^{\prime}}\bar{\sigma} \bigg ( S^{z}S^{\sigma}- S^{\sigma}S^{z} \bigg ) c^{\dagger}_{\vec{k}^{\prime} \bar{\sigma} \alpha}c^{}_{\vec{k} \sigma \alpha}\\
& \qquad = -\rho_{0} J_{\perp}J_{z}\frac{|\delta D|}{D} \sum_{\alpha} \sum_{\sigma} \sum_{\vec{k},\vec{k}^{\prime}}\sigma \bar{\sigma} S^{\sigma} c^{\dagger}_{\vec{k}^{\prime} \bar{\sigma} \alpha}c^{}_{\vec{k} \sigma \alpha}\\
& \qquad =\rho_{0} J_{\perp}J_{z}\frac{|\delta D|}{D}\sum_{\alpha} \sum_{\sigma} \sum_{\vec{k},\vec{k}^{\prime}} S^{\sigma} c^{\dagger}_{\vec{k}^{\prime} \bar{\sigma} \alpha}c^{\ }_{\vec{k} \sigma \alpha}
\end{split}
\end{equation*}
\begin{figure}[b]
\includegraphics[width=0.47\textwidth]{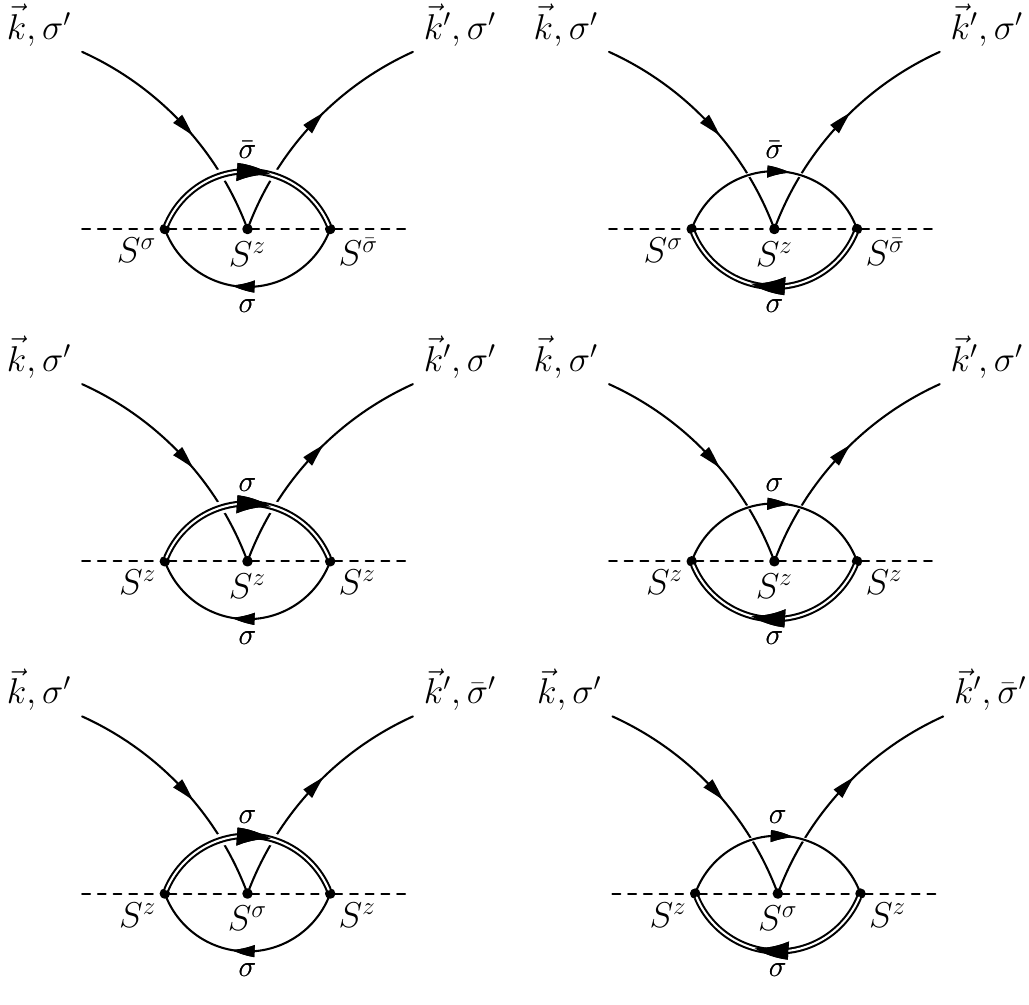}
\caption{Third-order poor man's scaling processes containing one fermionic loop. The diagram conventions are the same as for the second-order ones. Notice the edge-state fermion (double line) always connects the two outermost vertices, and the labels of the external fermion lines at the inner vertex enter the Hamiltonian as summation variables and can be relabeled interchangeably when collecting contributions. Here, again, the two diagrams in the first row contribute to the scaling of the $J_{z}$ coupling constant, while the other four diagrams contribute to the scaling of $J_{\perp}$.}
\label{fig:3rdPMD}
\end{figure}
Adding all the contributions from the bottom four diagrams in the figure, we get that they produce a rescaling of the perpendicular coupling constant given by
\begin{equation}
J_{\perp} \to J^{\prime}_{\perp} = J_{\perp}+2\rho_{0}\frac{|\delta D|}{D}J_{\perp}J_{z}
\end{equation}
This expression coincides with the previous one in the spin-isotropic case, with $J_{\perp}=J_{z}$. Notice the rescaling indicates that the effective couplings grow as the bandwidth is reduced (for the antiferromagnetic case).

\subsubsection{Third-order poor man's scaling}
In order to explore the existence of finite-coupling fixed points, at which the growing of the coupling constants stops, one needs to go beyond the leading order in the computation of the flow upon scaling. For the case of the Kondo model, that means considering the third-order corrections. As we shall see, the numerical prefactor in these corrections depends on the number of electronic channels interacting with the impurity in a degenerate way (the number of values, $K$, that the $\alpha$ index takes). For concreteness and easier comparison, we will assume the two-channel case during the derivation and later switch to the general result at the end.

To calculate third-order contributions to scaling, we proceed as before. But instead of approximating $H_{22}$ and $H_{00}$ with $H_{0}$, we approximate them with $H_{0}+H_{K}$, where $H_{K}$ is the Kondo interaction, and then we expand ${1}/(E-H_{22/00})$ up to the first order in $H_{K}$.
[Going to even higher order, one cannot neglect any longer $H_{02}$ and $H_{20}$ in the derivation of the effective Hamiltonian and has to revisit the derivation of Eq.~(\ref{eq:PMscalingH}) to include those contributions.]
The relevant corresponding diagrams contributing to the third-order flow of the coupling constant are those shown in Fig.~\ref{fig:3rdPMD}. The terms in the effective Hamiltonian connected to the top-left diagram (with a ``particle'' in the intermediate state) and the top-right diagram (with a ``hole'' in the intermediate state) are given by

\begin{widetext}
\small
\begin{equation*}
\begin{split}
& J^{2}_{\perp}J_{z} \sum\limits_{\alpha,\alpha^{\prime},\alpha^{\prime \prime}} \sum\limits_{\sigma,\sigma^{\prime},\sigma^{\prime \prime}}\sum\limits_{\vec{q}_{1},\vec{k}_{1}}\sum\limits_{\vec{k},\vec{k}^{\prime}}\sum\limits_{\vec{k}_{2},\vec{q}_{2}} \sigma S^{\sigma^{\prime}}S^{z}S^{\sigma^{\prime \prime}} c^{\dagger}_{\vec{k}_{1} \sigma^{\prime} \alpha^{\prime}}c^{\ }_{\vec{q}_{1} \bar{\sigma}^{\prime} \alpha^{\prime}}\frac{1}{E-H_{0}}c^{\dagger}_{\vec{k} \sigma \alpha}c^{\ }_{\vec{k}^{\prime} \sigma \alpha}\frac{1}{E-H_{0}} c^{\dagger}_{\vec{q}_{2} \sigma^{\prime \prime} \alpha^{\prime \prime}} c^{\ }_{\vec{k}_{2} \bar{\sigma}^{\prime \prime} \alpha^{\prime \prime}}\\
& = J^{2}_{\perp}J_{z} \sum\limits_{\alpha,\alpha^{\prime},\alpha^{\prime \prime}} \sum\limits_{\sigma,\sigma^{\prime},\sigma^{\prime \prime}} \sum\limits_{\vec{q}_{1},\vec{k}_{1}}\sum\limits_{\vec{k},\vec{k}^{\prime}}\sum\limits_{\vec{k}_{2},\vec{q}_{2}} \sigma  S^{\sigma^{\prime}}S^{z}S^{\sigma^{\prime \prime}} c^{\dagger}_{\vec{k}_{1} \sigma^{\prime} \alpha^{\prime}}c^{\ }_{\vec{q}_{1} \bar{\sigma}^{\prime} \alpha^{\prime}}c^{\dagger}_{\vec{k}^{\prime} \sigma \alpha}c^{\ }_{\vec{k} \sigma \alpha} c^{\dagger}_{\vec{q}_{2} \sigma^{\prime \prime} \alpha^{\prime \prime}} c^{\ }_{\vec{k}_{2} \bar{\sigma}^{\prime \prime} \alpha^{\prime \prime}} \frac{1}{\big ( E-\epsilon_{\vec{k}^{\prime}}+\epsilon_{\vec{k}}+\epsilon_{\vec{k}_{2}}-\epsilon_{\vec{q}_{2}}\big )\big (E+\epsilon_{\vec{k}_{2}}-\epsilon_{\vec{q}_{2}}\big )}
\end{split}
\end{equation*}
\normalsize
\\
\small
\begin{equation*}
\begin{split}
& J^{2}_{\perp}J_{z} \sum_{\alpha \alpha^{\prime} \alpha^{\prime \prime}}\sum\limits_{\sigma,\sigma^{\prime},\sigma^{\prime \prime}}\sum\limits_{\vec{q}_{1},\vec{k}_{1}}\sum\limits_{\vec{k},\vec{k}^{\prime}}\sum\limits_{\vec{k}_{2},\vec{q}_{2}} \sigma S^{\sigma^{\prime}}S^{z}S^{\sigma^{\prime \prime}} c^{\dagger}_{\vec{q}_{1} \bar{\sigma}^{\prime} \alpha^{\prime}} c^{\ }_{\vec{k}_{1} \sigma^{\prime} \alpha^{\prime}}\frac{1}{E-H_{0}}c^{\dagger}_{\vec{k} \sigma \alpha}c^{\ }_{\vec{k} \sigma \alpha}\frac{1}{E-H_{0}} c^{\dagger}_{\vec{k}_{2} \bar{\sigma}^{\prime \prime} \alpha^{\prime \prime}}c^{\ }_{\vec{q}_{2}\sigma^{\prime \prime}\alpha^{\prime \prime}}\\
& = J^{2}_{\perp}J_{z} \sum_{\alpha \alpha^{\prime} \alpha^{\prime \prime}} \sum\limits_{\sigma,\sigma^{\prime},\sigma^{\prime \prime}}\sum\limits_{\vec{q}_{1},\vec{k}_{1}}\sum\limits_{\vec{k},\vec{k}^{\prime}}\sum\limits_{\vec{k}_{2},\vec{q}_{2}} \sigma S^{\sigma^{\prime}}S^{z}S^{\sigma^{\prime \prime}} c^{\dagger}_{\vec{q}_{1} \bar{\sigma}^{\prime} \alpha^{\prime}} c^{\ }_{\vec{k}_{1} \sigma^{\prime} \alpha^{\prime}} c^{\dagger}_{\vec{k}^{\prime} \sigma \alpha}c^{\ }_{\vec{k} \sigma \alpha} c^{\dagger}_{\vec{k}_{2} \bar{\sigma}^{\prime \prime} \alpha^{\prime \prime}}c^{\ }_{\vec{q}_{2}\sigma^{\prime \prime}\alpha^{\prime \prime}}\frac{1}{\big( E-\epsilon_{\vec{k}^{\prime}}+\epsilon_{\vec{k}}-\epsilon_{\vec{k}_{2}}+\epsilon_{\vec{q}_{2}}\big)\big( E-\epsilon_{\vec{k}_{2}}+\epsilon_{\vec{q}_{2}}\big)}
\end{split}
\end{equation*}
\normalsize \qquad \\
\end{widetext}

In the second line of the first expression above, we set $H_{0} = 0$, so that we are measuring energies from the ground state, as we did in the second-order calculation. The diagram contribution is obtained by contracting $\vec{k}_{1}$ and  $\vec{k}_{2}$ as well as the ``fast'' modes, $\vec{q}_{1}$ and  $\vec{q}_{2}$. In addition, we approximate $\epsilon_{\vec{q}} \approx D$ and $\sum_{\vec{q}} \rightarrow \rho_{0}|\delta D|$, and we assume $\epsilon_{\vec{k}},\ \epsilon_{\vec{k}{\prime}}, \ E << D$, so that the expression for the diagram reduces to
\begin{equation*}
\begin{split}
J^{2}_{\perp}J_{z} \sum_{\alpha, \alpha^{\prime}}\sum_{\sigma, \sigma^{\prime}}\sum\limits_{\vec{k},\vec{k}^{\prime}}&  \sigma S^{\sigma^{\prime}}S^{z}S^{\bar{\sigma}^{\prime}}c^{\dagger}_{\vec{k} \sigma \alpha}c^{}_{\vec{k}^{\prime} \sigma \alpha} \sum\limits_{\vec{k}_{1}}\frac{\rho_{0} |\delta D|}{\big ( \epsilon_{\vec{k}_{1}}-D\big )^2}
\end{split}
\end{equation*}

We can go from the summation in $\vec{k}_{1}$ to an integration in energy by using $\sum_{\vec{k}_{1}} \to \int\limits_{-D + |\delta D|}^{0} \rho_{0}\,d\epsilon_{1}$. Where we are integrating over all the filled nonedge states, (here $\rho_{0}$ depends on the dimensionality. and we take it to be a constant throughout the calculation; in the more general case it can have an energy dependence that contributes to the integral). After performing the integration we arrive at the expression
\begin{equation}
\begin{split}
\rho^{2}_{0} J^{2}_{\perp} J_{z}\frac{|\delta D|}{2D} \sum_{\alpha, \alpha^{\prime}} \sum_{\sigma, \sigma^{\prime}} \sum\limits_{\vec{k},\vec{k}^{\prime}} \sigma S^{\sigma^{\prime}}S^{z}S^{\bar{\sigma}^{\prime}}c^{\dagger}_{\vec{k}^{\prime} \sigma \alpha}c^{}_{\vec{k} \sigma \alpha}
\end{split}
\end{equation}

\begin{figure}[b]
\includegraphics[width=0.47\textwidth]{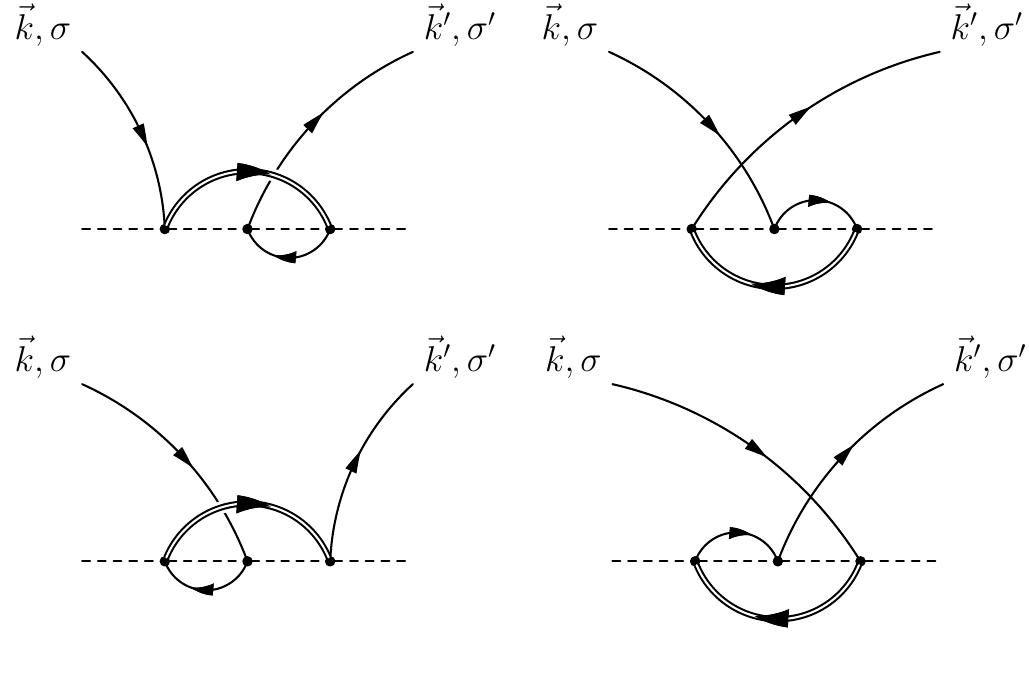}
\caption{Third-order diagrams without fermionic loops. The diagrams in the first row are canceled partially by the ones in the second. The remaining contribution does not, however, contribute to the flow of the coupling constant since it belongs to a class of ``reducible'' diagrams.}
\label{fig:nonloop}
\end{figure}

Using the spin-1/2 facts that $S^{\bar{\sigma}^{\prime}}S_{z} = -\frac{1}{2}\bar{\sigma}^{\prime} S^{\bar{\sigma}^{\prime}}$ and  $S^{\bar{\sigma}^{\prime}}S^{\sigma^{\prime}} = \frac{1}{2}+\bar{\sigma}^{\prime}S_{z}$, we arrive at
\begin{equation*}
\begin{split}
 -\rho^{2}_{0} J^{2}_{\perp} J_{z}\frac{|\delta D|}{2D} \sum_{\alpha, \alpha^{\prime}} \sum_{\sigma, \sigma^{\prime}} \sum\limits_{\vec{k},\vec{k}^{\prime}} \sigma \bigg(\frac{\sigma^{\prime}}{4} + \frac{1}{2}S^{z} \bigg)c^{\dagger}_{\vec{k}^{\prime} \sigma \alpha}c^{}_{\vec{k} \sigma \alpha}
\end{split}
\end{equation*}
In a similar way as in the case of the particle diagram, for the hole diagram we get a contribution of the form
\begin{equation*}
\begin{split}
 -\rho^{2}_{0} J^{2}_{\perp} J_{z}\frac{|\delta D|}{2D} \sum_{\alpha, \alpha^{\prime}} \sum_{\sigma, \sigma^{\prime}} \sum\limits_{\vec{k},\vec{k}^{\prime}} \sigma \bigg(\frac{\sigma^{\prime}}{4} + \frac{1}{2}S^{z} \bigg)c^{\dagger}_{\vec{k}^{\prime} \sigma \alpha}c^{\ }_{\vec{k} \sigma \alpha}
\end{split}
\end{equation*}
Together with the particle contribution, we get a potential scattering term, (which we disregard, since it is too high order to contribute to the wave-function renormalization that will be discussed below), and a term that contributes to the rescaling of the $J_{z}$ coupling constant
\begin{equation}
\begin{split}
 -\rho^{2}_{0} J^{2}_{\perp} J_{z}\frac{|\delta D|}{2D} \sum_{\alpha, \alpha^{\prime}} \sum_{\sigma, \sigma^{\prime}} \sum\limits_{\vec{k},\vec{k}^{\prime}} \sigma S^{z}c^{\dagger}_{\vec{k}^{\prime} \sigma \alpha}c^{\ }_{\vec{k} \sigma \alpha}
\end{split}
\end{equation}
After summing over the free spin and channel indices, ($\sigma^{\prime}$ contributing a factor of $2$ and $\alpha^{\prime}$ contributing a factor of $K=2$, respectively), we arrive at the final contribution to the flow of the parallel coupling constant
\begin{equation}
\begin{split}
 -2 \rho^{2}_{0} J^{2}_{\perp} J_{z}\frac{|\delta D|}{D} \sum_{\alpha} \sum_{\sigma} \sum\limits_{\vec{k},\vec{k}^{\prime}} \sigma S^{z}c^{\dagger}_{\vec{k}^{\prime} \sigma \alpha}c^{\ }_{\vec{k} \sigma \alpha}
\end{split}
\end{equation}
The same procedure can be repeated for all the remaining diagrams in Fig.~\ref{fig:3rdPMD}. In that way one gets the scaling of the coupling constants, up to third order, to apparently be
\begin{equation}\label{scalingI}
\begin{split}
& J^{\prime}_{z} \simeq J_{z}+2\rho_{0}\frac{|\delta D|}{D}J^{2}_{\perp}- 2\rho^{2}_{0}\frac{|\delta D|}{D}J^{2}_{\perp}J_{z}+ \rho^2_{0} \frac{|\delta D|}{D}J^{3}_{z} \\
& J^{\prime}_{\perp} \simeq J_{\perp}+ 2\rho_{0}\frac{|\delta D|}{D} J_{\perp}J_{z}-\rho^{2}_{0}\frac{|\delta D|}{D}J_{\perp} J^{2}_{z}
\end{split}
\end{equation}
This includes only the contribution coming from the diagrams in Fig.~\ref{fig:3rdPMD}, the ones with a fermionic loop that counts the number of channels. There are, however, other diagrams one can make that do not have a fermionic loop, and might seem to give a channel-independent contribution to the third-order flow of the coupling constant. Such diagrams are shown schematically in Fig.~\ref{fig:nonloop}. As it turns out, these diagrams do not contribute to the scaling of the coupling constant, because they belong to the class of ``reducible'' diagrams \cite{kuramoto1998} which can be created by combining lower-order contributions (see Ref.~\onlinecite{Solyom1974} for an explicit calculation including these diagrams in the framework of field-theoretic scaling and renormalization). Namely, diagrams in Fig.~\ref{fig:nonloop} can be regarded as arising from contracting an additional Kondo vertex to one of the second-order diagrams shown in Fig.~\ref{fig:2ndPMD}. (Or, alternatively, as taking one of those second-order diagrams and replacing one of its vertices by a renormalized one also taken from the same set of second-order diagrams.) As such, these diagrams do not contribute to the further independent scaling of the Kondo coupling. There are, nevertheless, other corrections to the flow that are still unaccounted for. We shall turn to those next.

\subsubsection{Wave-function renormalization}

At this point we need to pause and look back at the potential-scattering contributions that we obtained while integrating out the band-edge states. We shall see that they contribute as well to the rescaling of the coupling constant, although indirectly. This effect is part of the so-called \textit{wave-function renormalization}.

All those terms that contributed to potential scattering in the second-order calculation, (and that we have so far neglected), need to be taken into account if we are going to consider third-order impurity-scattering processes. The first potential-scattering terms that we came across were from the two topmost diagrams in Fig.~\ref{fig:2ndPMD}, and were given by Eqs.~(\ref{potentialP}) and (\ref{eq:potentialH}), respectively.
We are now interested in the (relative) ground-state energy shift that is induced by these contributions. So we set $\vec{k}^{\prime}=\vec{k}$ and replace the sums over $\vec{k}$ for integrals in energy,
\begin{equation}\label{eq:A17}
\begin{split}
&\frac{1}{2}J^{2}_{\perp} \rho^{2}_{0}|\delta D|  \sum_{\alpha} \sum_{\sigma}\int\limits_{-D+|\delta D|}^{0}\frac{d\epsilon}{(E-D+\epsilon)}\\
&\frac{1}{2}J^{2}_{\perp} \rho^{2}_{0}|\delta D|  \sum_{\alpha} \sum_{\sigma}\int\limits_{0}^{D-|\delta D|}\frac{d\epsilon}{(E-D-\epsilon)}
\end{split}
\end{equation}
Notice the difference in the integration limits between particle and hole diagrams. This is because in the particle diagram we are integrating over the (filled) particle states and in the hole diagram over the (empty-particle) hole states. One can readily see that both of the integrals turn out to be exactly the same. We add them up and carry them out to obtain
\begin{equation}
\begin{split}
 &J^{2}_{\perp} \rho^{2}_{0} |\delta D| \sum_{\alpha} \sum_{\sigma} \ln \bigg( \frac{E-D}{E-2D} \bigg )\\
 & \approx -4 J^{2}_{\perp} \rho^{2}_{0} |\delta D|\bigg( \ln(2) -\frac{E}{2D} \bigg )\\
\end{split}
\end{equation}
where, going into the second line, we have also done the summations in $\sigma$ and $\alpha$, as well as expanded the logarithm and kept only terms that are linear in $E$.

\begin{figure}[t]
\includegraphics[width=0.47\textwidth]{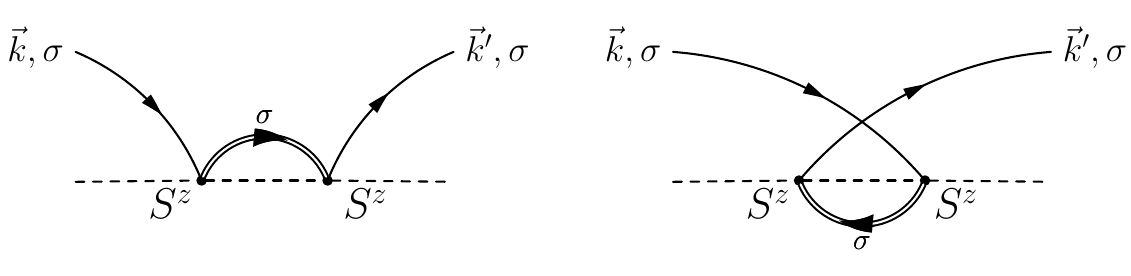}
\caption{Second-order diagrams contributing to potential scattering.}
\label{fig:extraJz}
\end{figure}

Additional potential-scattering terms that we have not yet considered are shown in Fig.~\ref{fig:extraJz}. These are processes involving two $S_{z}$ vertices. Doing the edge-mode integration, which is done in a same way as for any other second-order diagram, one arrives at
\begin{equation}
\frac{1}{4}J^{2}_{z}\rho_{0}|\delta D| \sum_{\alpha} \sum_{\sigma} \sum_{\vec{k},\vec{k}^{\prime}} c^{\dagger}_{\vec{k}^{\prime} \sigma \alpha}c_{\vec{k} \sigma \alpha}\frac{1}{(E-D+\epsilon_{\vec{k}})}
\end{equation}
As in the previous cases, calculating the expectation value that gives the shift in the ground-state energy, (with $\vec{k}^{\prime}=\vec{k}$), and taking into account that both, particle and hole, diagrams give the same integral in energy, one gets
\begin{equation}
\begin{split}
&\frac{1}{2}J^{2}_{z}\rho^{2}_{0}|\delta D|\sum_{\alpha} \sum_{\sigma} \ln \bigg ( \frac{E-D}{E-2D} \bigg )\\
&\approx -2 J^{2}_{z}\rho^{2}_{0}|\delta D| \bigg (\ln(2)+\frac{E}{2D} \bigg )
\end{split}
\end{equation}
Adding the two energy-shift contributions we finally arrive at the total shift being
\begin{equation}\label{eq:eshiftD}
-\bigg (\ln(2)+\frac{E}{2D} \bigg ) \bigg (2 J^{2}_{z}\rho^2_{0}|\delta D| +4J^{2}_{\perp}\rho^{2}_{0}|\delta D| \bigg )
\end{equation}

The ground-state energy shift coming from the third-order diagrams can be disregarded, because it will contribute to a higher order in the flow equations. These energy shifts are new terms in our Hamiltonian introduced by the scaling procedure. Basically, after the scaling procedure we move away from our original Kondo Hamiltonian to a new one which we can write as
\begin{equation*}
\begin{split}
H^{\prime} - \ln(2) \bigg (2 J^{2}_{z}\rho_{0}|\delta D| & +4J^{2}_{\perp}\rho^{2}_{0}|\delta D| \bigg )\\
& -E\bigg ( J^{2}_{z}\rho_{0}\frac{|\delta D|}{D} +2J^{2}_{\perp}\rho^{2}_{0}\frac{|\delta D|}{D} \bigg )
\end{split}
\end{equation*}
where $H^{\prime}$ has the same form as the original Hamiltonian (free electrons plus Kondo interaction), with coupling constants $J^{\prime}_{\perp}$ and $J^{\prime}_{z}$ given by the scaling laws in Eq.~(\ref{scalingI}). We see that our ``new'' Hamiltonian is energy dependent through the second term in the above expression. This energy dependence can be removed by solving the secular equation
\begin{equation*}
\begin{split}
E = & \frac{1}{\bigg( 1+ J^{2}_{z}\rho_{0}\frac{|\delta D|}{D} +2 J^{2}_{\perp}\rho^{2}_{0}\frac{|\delta D|}{D}\bigg)}H^{\prime}\\
&- \ln(2)\frac{\bigg ( 2J^{2}_{z}\rho_{0}|\delta D|+4J^{2}_{\perp}\rho^{2}_{0}|\delta D| \bigg )}{\bigg( 1+ J^{2}_{z}\rho_{0}\frac{|\delta D|}{D} +2J^{2}_{\perp}\rho^{2}_{0}\frac{|\delta D|}{D}\bigg)}
\end{split}
\end{equation*}
The second term is a constant that gives a uniform shift in energy. We can simply disregard it, since we are free to choose the zero of energy. That way we get that the energy-independent effective Hamiltonian is given by
\begin{equation}
 \bigg( 1- J^{2}_{z}\rho_{0}\frac{|\delta D|}{D} -2J^{2}_{\perp}\rho^{2}_{0}\frac{|\delta D|}{D}\bigg)H^{\prime} 
\end{equation}
To third order, this gives the following modified scaling laws for the coupling constants:
\begin{equation}
\begin{split}
& J^{\prime}_{\perp} = J_{\perp}+ 2\rho_{0}\frac{|\delta D|}{D} J_{\perp}J_{z}-2\rho^{2}_{0}\frac{|\delta D|}{D}J^{2}_{z}J_{\perp}-2\rho^{2}_{0}\frac{|\delta D|}{D}J^{3}_{\perp} \\
& J^{\prime}_{z} = J_{z}+2\rho_{0}\frac{|\delta D|}{D}J^{2}_{\perp}- 4 \rho^{2}_{0}\frac{|\delta D|}{D}J^{2}_{\perp}J_{z}
\end{split}
\end{equation}
Notice the $J_{z}^3$ term disappeared from the flow of $J_{z}$, while a $J_{\perp}^3$ term appeared in the flow of $J_{\perp}$. 
One can now redefine the coupling constants by absorbing the density of states, $g_{\perp} = 2\rho_{0}J_{\perp}$ and $g_{z} = 2\rho_{0}J_{z}$, to arrive at the following expressions for the so-called $\beta$ functions:
\begin{equation}
\begin{split}
& \frac{d g_{\perp}}{d\ln(D)} = -g_{\perp}g_{z} + \frac{M}{2} g_{\perp}g_{z}^{2} + \frac{M}{2} g_{\perp}^{3}\\
& \frac{d g_{z}}{d\ln(D)} = -g_{\perp}^{2} + Mg_{z}g_{\perp}^{2}
\end{split}
\end{equation}
where we included the dependence on the number of channels, $K=2M$. (We had been assuming $M=1$, or the two-channel case, but it is straightforward to check that one simply needs to insert a factor of $M$ in all the cubic terms and the resulting expression is valid for arbitrary $K\in\mathbb{N}$.) As we will see later, these functions capture important information about the low-energy physics of the (multichannel) Kondo model, so we are interested in knowing if the compactification procedure is able to capture it.

\subsection{Conventionally compactified model}

After a mapping to 1D and focusing again in the two-channel case (to keep the notation simpler; one can reinsert the generic value of $M$ at the end), the \textit{conventional} refermionization \cite{Ljepoja2024a} of the spin-flip interaction is given by
\begin{equation}\label{eq:flipC}
\begin{split}
H_{K}^{\perp} & = -J_{\perp} S^{-} \sum_{k,k^{\prime}}c^{\dagger}_{k,s} c_{k^{\prime},sl}-J_{\perp} S^{+}\sum_{k,k^{\prime}} c^{\dagger}_{k,sl}c_{k^{\prime},s}\\
&\qquad + J_{\perp} S^{-}\sum_{k,k^{\prime}}c_{k,sl}^{\dagger}c_{k^{\prime},s}^{\dagger}- J_{\perp} S^{+}\sum_{k,k^{\prime}} c_{k,sl}^{}c_{k^{\prime},s}^{}
\end{split}
\end{equation}
while, at the same time, the parallel (or Ising) part of the Kondo interaction is given by
\begin{equation}\label{eq:parallelC}
\begin{split}
H^{z}_{K}& = 2S^{z} J_{z}\sum_{k, k^{\prime}}c_{k,s}^{\dagger}c_{k^{\prime},s}
\end{split}
\end{equation}

As we did for the direct-model calculation, we can single out the parts of Hamiltonian that create a particle in the upper band edge,
\begin{equation*}
\begin{split}
H_{21}^{\perp} & = -J_{\perp} S^{-} \sum_{k,q}c^{\dagger}_{q,s} c_{k,sl}- J_{\perp} S^{+}\sum_{k,q} c^{\dagger}_{q,sl}c_{k,s}\\
&\qquad + J_{\perp} S^{-}\sum_{k,q}c_{q,sl}^{\dagger}c_{k,s}^{\dagger}+ J_{\perp} S^{-}\sum_{k,q}c_{k,sl}^{\dagger}c_{q,s}^{\dagger}
\end{split}
\end{equation*}
and
\begin{equation*}
\begin{split}
H^{z}_{21}& = 2S^{z} J_{z}\sum_{k,q}c_{q,s}^{\dagger}c_{k,s}
\end{split}
\end{equation*}
Likewise, we can also specify the parts of the Hamiltonian that create a hole in the lower band edge, (these can be obtained from the above by swapping $k$ and $q$ and performing hermitian conjugation as needed),
\begin{equation*}
\begin{split}
H_{01}^{\perp} & = -J_{\perp} S^{-} \sum_{k, q}c^{\dagger}_{k,s}c_{q,sl} -J_{\perp} S^{+}\sum_{k,q} c^{\dagger}_{k,sl}c_{q,s}\\
&\qquad - J_{\perp} S^{+}\sum_{k,q} c_{q,sl}^{}c_{k,s}^{}
-J_{\perp} S^{+}\sum_{k,q} c_{k,sl}^{}c_{q,s}^{}
\end{split}
\end{equation*}
and
\begin{equation*}
\begin{split}
H^{z}_{01}& = 2S^{z} J_{z}\sum_{k, q}c_{k,s}^{\dagger}c_{q,s}
\end{split}
\end{equation*}

These terms and their Hermitian conjugates will be used in the effective Hamiltonian of Eq.~(\ref{eq:PMscalingH}). The ensuing calculation proceeds similarly as before. Notice, however, that in the case of the conventionally compactified model we have interaction vertices that either create or destroy a pair of fermionic degrees of freedom at the same time. These are what we will call \textit{anomalous} vertices, to distinguish them from \textit{normal} ones, which (at the same time) create one fermion and destroy another while conserving their total number.

\subsubsection{Second-order scaling}

\begin{figure}[h!]
\includegraphics[width=0.49\textwidth]{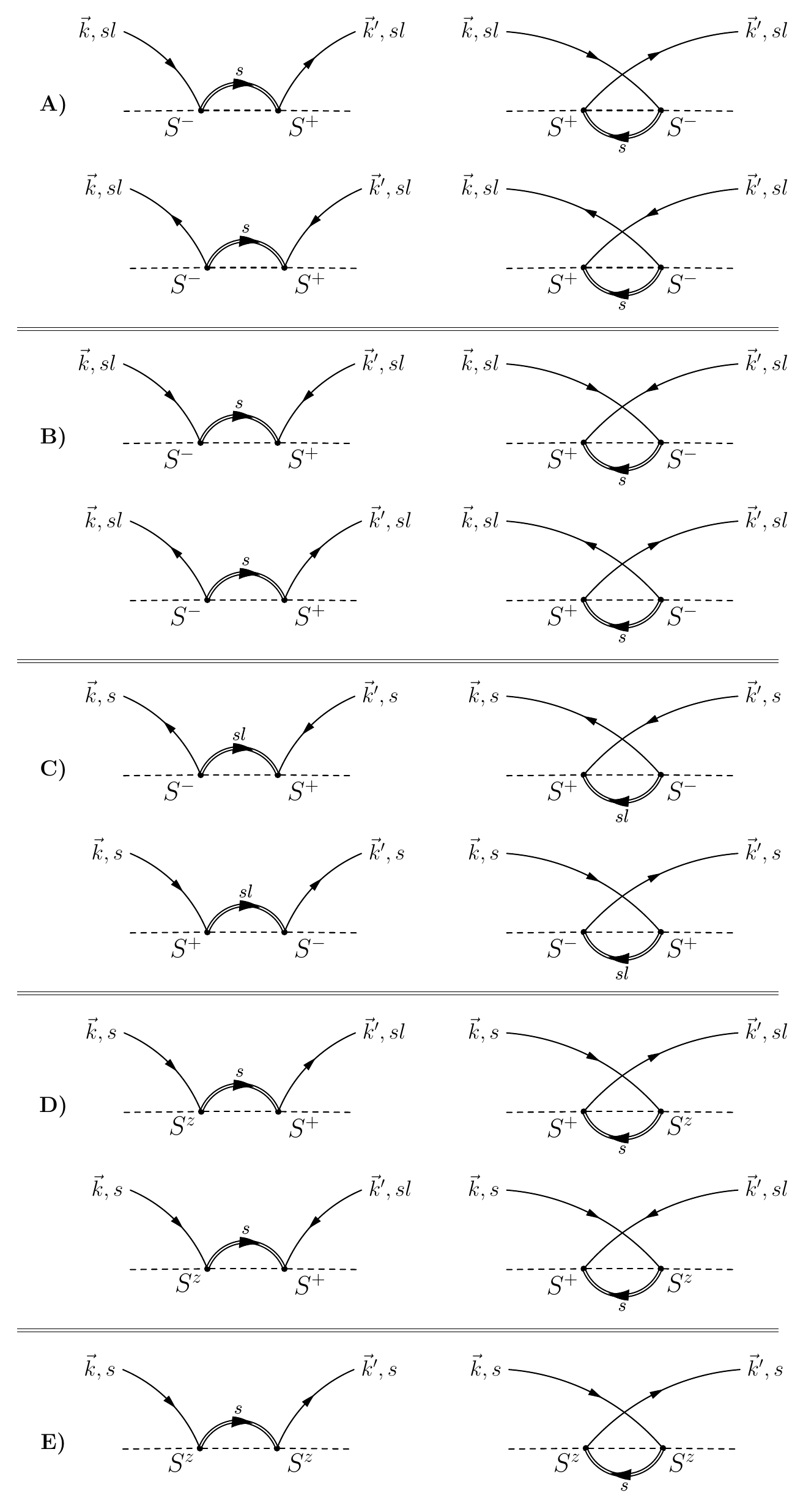}
\caption{Second-order poor man's scaling processes. The dashed line represents the impurity, full lines are inner-band fermions, and the double line is the scattered edge-state fermion. They are separated into five groups based on their contributions. Diagrams in group (A) cancel amongst each other. Diagrams in group (B) are unphysical (``untranslatable''), and they give a new vertex in the Hamiltonian which is zero when we keep the calculation to order ${1}/{D}$. Diagrams in groups (C) and (D) contribute to the scaling of $J_{\perp}$ and $J_{z}$, respectively. Diagrams in group (E) contribute to potential scattering, which will affect the wave-function renormalization.}
\label{fig:2ndPMC}
\end{figure}

Processes contributing to the coupling-constant flow up to the second order in the \textit{conventional} compactification of the Kondo model are given in Fig.~\ref{fig:2ndPMC}. They are separated into groups depending on the different ways they contribute to the scaling.
The first group of diagrams, labeled (A), will cancel each other and their combined contribution will be zero. Therefore, there will be no contribution to parallel scattering generated in the $sl$ sector. One can see how the cancellation happens by the fact that they have the same internal structure while the external fermionic-leg arrows are reversed between the two rows. (This reversal introduces a relative minus sign and hence the cancellation.) 

In group (B) we have the diagrams that are made up by combining \textit{normal} and \textit{anomalous} vertices. This is an important characteristic of the conventionally refermionized model, that it allows for such mixing of normal and anomalous vertices in a diagram. And it is this mixing of the \textit{normal} and \textit{anomalous} vertices that is responsible for the appearance of ``unphysical'' diagrams which are impossible to translate into the language of the original, direct-model, fermions.
\begin{figure}[h]
\includegraphics[width=0.49\textwidth]{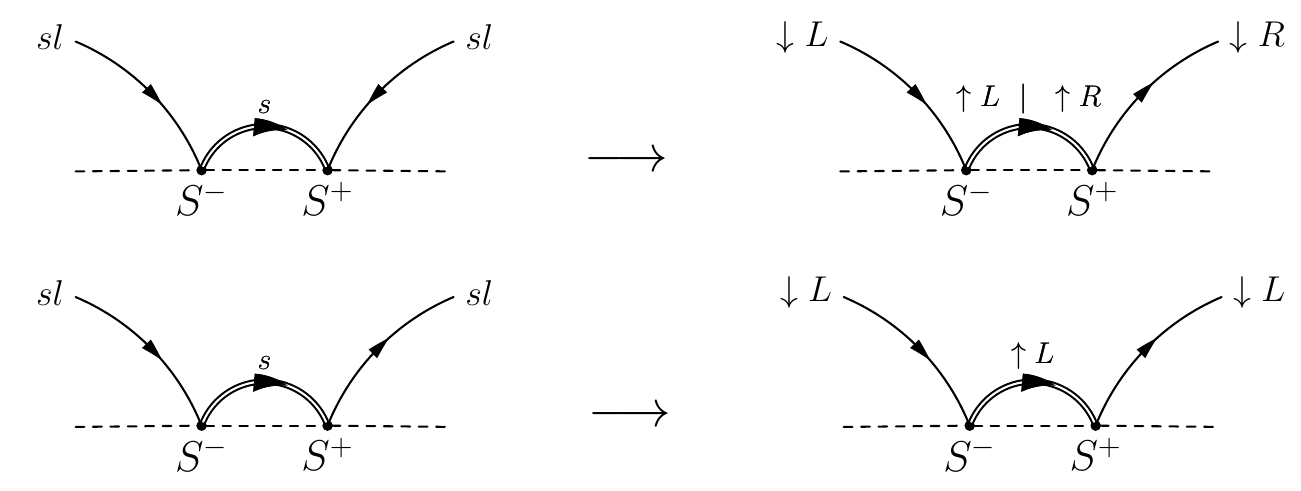}
\caption{Examples of translation of conventional-model diagrams into the direct-model language. The first row shows that a conventional diagram that mixes \textit{normal} and \textit{anomalous} vertices would translate into a process that needs to flip from $L$ to $R$ ``on the fly'', along a single fermion propagator (which is not allowed by $H_{0}$). This is an example of an unphysical contribution to the scaling taken from group (B). On the other hand, the second row shows how a very similar diagram, taken from group (A), but with two \textit{normal} vertices translates without problems (and the same would be true, \textit{mutatis mutandis}, for the corresponding diagram with two \textit{anomalous} vertices).}
\label{fig:unph}
\end{figure}
Combining them we get an expression that introduces a new term to our original Hamiltonian which takes the form of an anomalous potential scattering (\textit{i.e.}, scattering independent of the impurity spin)
\begin{equation}\label{eq:unphy}
\begin{split}
&J^{2}_{\perp}\sum_{k,k^{\prime}} c_{k,sl}c_{k^{\prime},sl}\frac{\rho_{0}|\delta D| }{(E-D + \epsilon_{k^{\prime}})}\\
& \qquad +J^{2}_{\perp} \sum_{k,k^{\prime}} c^{\dagger}_{k,sl}c^{\dagger}_{k^{\prime},sl}\frac{\rho_{0}|\delta D| }{(E-D - \epsilon_{k^{\prime}})}
\end{split}
\end{equation}
Such a term has no correlate in the original Kondo Hamiltonian and has the form of a fermionic pair creation/annihilation operator. However, in the limit when $D \gg E, \epsilon_{k}$ and if we are keeping only $1/D$ contributions, this processes are zero (being the Fourier transform of the product of two fermionic creation/annihilation operators at the same real-space location, the site of the impurity).

On the other hand, diagrams in group (C) of Fig.~\ref{fig:2ndPMC} contribute to the renormalization of the $J_{z}$ coupling constant. Their combined contribution is
\begin{equation}
\begin{split}
4 J^{2}_{\perp}\rho_{0} \frac{|\delta D|}{D} S^{z}\sum_{k,k^{\prime}} c^{\dagger}_{k,s}c_{k^{\prime},s}
\end{split}
\end{equation}
Comparing this term with the parallel interaction in the conventionally compactified Hamiltonian, Eq.~(\ref{eq:parallelC}), we arrive at the same flow as we obtained in the direct-model calculation.

Meanwhile, the diagrams that contribute to the scaling of the perpendicular coupling constant, $J_{\perp}$, are shown as group (D) in Fig.~\ref{fig:2ndPMC}. Their combined contribution to the flow is
\begin{equation}
\begin{split}
& - 2 J_{\perp}J_{z}\rho_{0} \frac{|\delta D|}{D} S^{+} \sum_{k,k^{\prime}}c^{\dagger}_{k,sl}c_{k^{\prime},s}\\
& \qquad - 2 J_{\perp}J_{z} \rho_{0} \frac{|\delta D|}{D} S^{+}\sum_{k,k^{\prime}}c_{k,sl}c_{k^{\prime},s}
\end{split}
\end{equation}
As was the case with the parallel interaction, the flow that we obtain using the conventionally compactified model is the same as in the direct calculation.
In addition to the flow of $J_{\perp}$, these diagrams also contribute to potential scattering as 
\begin{equation}\label{eq:wfrC1}
\begin{split}
& J^{2}_{\perp}\sum_{k,k^{\prime}}\bigg ( c^{\dagger}_{k,s}c_{k^{\prime},s} +c^{\dagger}_{k,sl}c_{k^{\prime},sl} \bigg )\frac{\rho_{0}|\delta D| }{(E-D + \epsilon_{k^{\prime}})}\\
& + J^{2}_{\perp} \sum_{k,k^{\prime}}\bigg ( c_{k,s}c^{\dagger}_{k^{\prime},s}+ c_{k,sl}c^{\dagger}_{k^{\prime},sl} \bigg ) \frac{ \rho_{0}|\delta D| }{(E-D - \epsilon_{k^{\prime}})}
\end{split}
\end{equation}
These contributions will be dealt with in the same way as in the direct calculation (by incorporating the effects of the associated wave-function renormalization). 

Finally, the last two diagrams in the Fig.~\ref{fig:2ndPMC}, the ones in group (E), contribute solely to the potential scattering leading to wave-function renormalization. They give
\begin{equation}\label{eq:wfrC2}
\begin{split}
 & J^{2}_{z} \sum_{k,k^{\prime}}c^{\dagger}_{k,s}c_{k^{\prime},s}\frac{ \rho_{0}|\delta D| }{(E-D + \epsilon_{k^{\prime}})}\\
 & \quad +J^{2}_{z} \sum_{k,k^{\prime}}c_{k,s}c^{\dagger}_{k^{\prime},s}\frac{ \rho_{0}|\delta D| }{(E-D - \epsilon_{k^{\prime}})}
\end{split}
\end{equation}

Looking at all the diagrammatic contributions to the second-order scaling of the conventionally refermionized model, one sees that we arrive at the same flow as in the direct calculation. Even though, as shown in Fig.~\ref{fig:unph}, there are diagrams that are unphysical. It just so happens that they do not contribute to the scaling of the Kondo interaction and, furthermore, this unphysical contribution disappears in the limit of a large cutoff. As in the direct-model calculation, we proceed further to look at the third-order contributions.

\subsubsection{Third-order scaling}
\begin{figure}[h]
\includegraphics[width=0.47\textwidth]{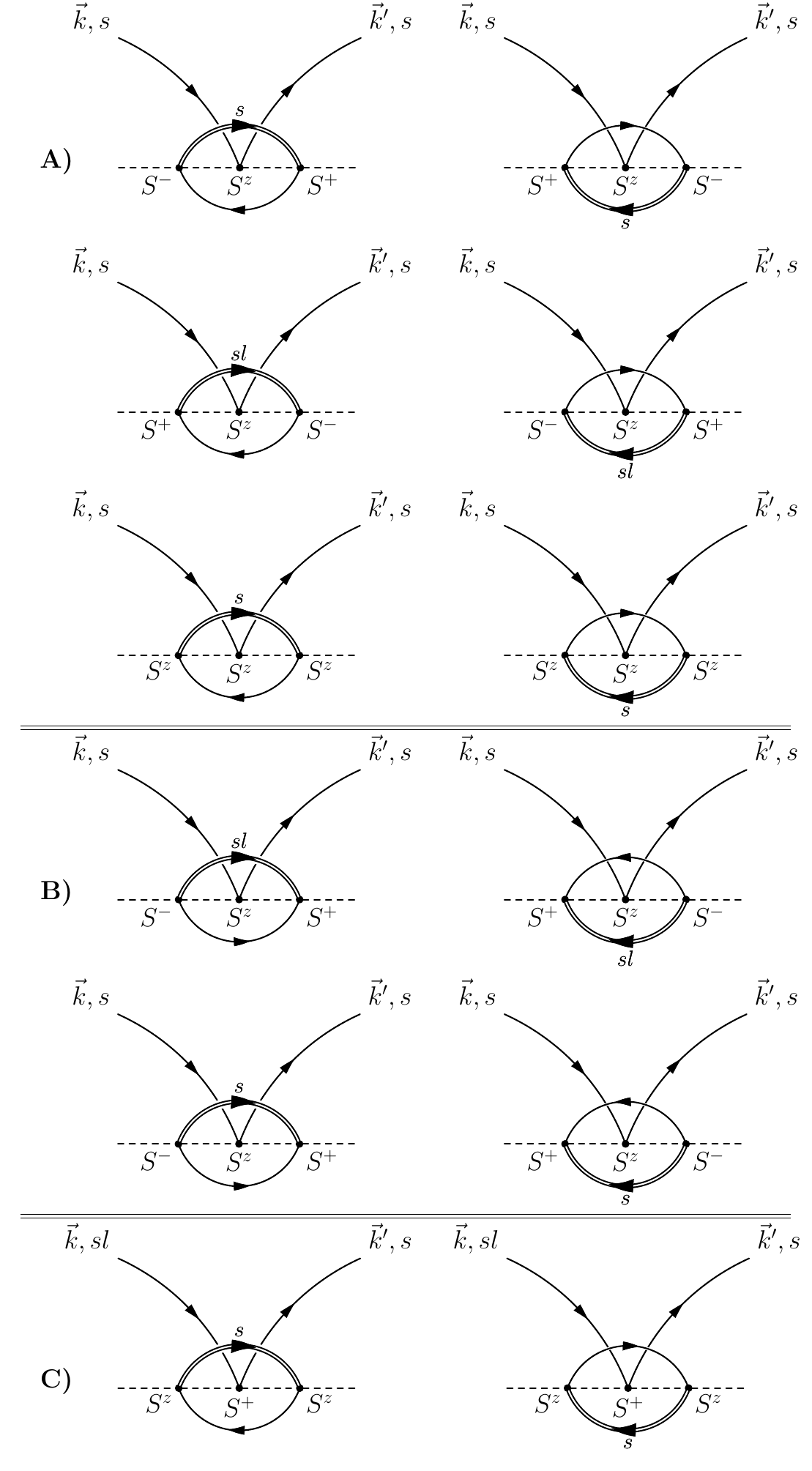}
\caption{Third-order poor man's scaling diagrams for the \textit{conventionally} compactified model. The dashed line represents the impurity, full lines are inner-band fermions, and the double line is a scattered edge-state fermion with momentum $\vec{q}$. Diagrams in group (A) those that contribute to the re-scaling of the parallel coupling constant where all the vertices are normal. Diagrams in group (B) also contribute to the scaling of $J_{z}$, but involve two anomalous spin-flip vertices. Finally, the set of diagrams in group (C) contributes to the scaling of the $J_{\perp}$ coupling.}
\label{fig:3rdPMC}
\end{figure}

The relevant third-order T-matrix diagrams are shown in Fig.~\ref{fig:3rdPMC}.
Diagrams in group (A) contain only normal vertices and contribute to the flow of the parallel coupling constant. Notice we have two different contributions to the flow, they are the $J^{2}_{\perp}J_{z}$ and the $J^{3}_{z}$ contributions. Additionally, in group (B) we have diagrams that mix normal and anomalous vertices, (the middle vertex is normal while the ones on the left and right are anomalous), and also contribute to the flow of the parallel coupling constant. It is important to note that the diagrams in group (B), although mixing normal and anomalous vertices, are not themselves unphysical. Meaning, they have a valid translation into the direct-model language. The reason for that is that the two outer vertices on the left and right in these third-order diagrams have to be either both anomalous [as in group (B)] or both normal [as in group (A)], since they have to be able to form a fermion loop. In turn, that means that along the fermion loop there will not be mixing of vertices and that the loop  will always be translatable. As far as the middle vertex is concerned, it is always translatable also since it makes no contractions, it is just a free, open vertex. This goes further than just for the flow of the  $J_{z}$ coupling constant. Since, as we saw in the previous section, no/open fermion-loop diagrams (the ones shown in Fig.~\ref{fig:nonloop}) are not contributing, every third-order diagram one makes in the conventional language will be translatable in terms of the direct-model fermions, and therefore physical. 

The calculation of the contributions from these diagrams proceeds in the same way as it was done for the direct model. Contractions are done in the same way weather we are contracting anomalous or normal vertices in the fermion loop (they are just one reversal of the fermionic-propagator direction away from each other). Carrying out the calculation for all the diagrams that contribute to the flow of the parallel coupling constant we arrive at their full combined contribution being
\begin{equation}
\begin{split}
& 2 J^{3}_{z} S^{z}\frac{\rho^{2}_{0}|\delta D| }{D}\sum_{k,k^{\prime}} c^{\dagger}_{k,s}c_{k^{\prime},s}\\
& \qquad - 4 \rho^{2}_{0} J^{2}_{\perp} J_{z}S^{z}\frac{|\delta D| }{D}\sum_{k,k^{\prime}} c^{\dagger}_{k,s}c_{k^{\prime},s}
\end{split}
\end{equation}
This is the same contribution as in the direct case, which is as expected considering that there are no unphysical contributions in these third-order diagrams. In addition to the spin-scattering contributions, there are also potential-scattering terms that were disregarded here since they will only influence the flow at the fourth order, (which is higher than what we are aiming for and results practical within a poor man's framework). 

Diagrams in group (C) of Fig.~\ref{fig:3rdPMC} contribute to the rescaling of the perpendicular coupling constant. For these diagrams, in the same way as for the diagrams in group (B), mixing of the anomalous and normal vertices is physical and translatable. In this case, however, only the middle vertex can be either anomalous or normal, while the ones on the left and right are always normal. Once again, their contribution is calculated using the same steps we used to calculate the direct-model diagrams, and we arrive at
\begin{equation*}
\begin{split}
J^{2}_{z} J_{\perp}S^{+}\frac{\rho^{2}_{0}|\delta D| }{D}\sum_{k,k^{\prime}}c^{\dagger}_{k^{\prime},s}c_{k,sl}\\
\end{split}
\end{equation*}
which is, as expected, the same contribution we had in the direct scaling scheme.

\begin{figure}[h]
\includegraphics[width=0.47\textwidth]{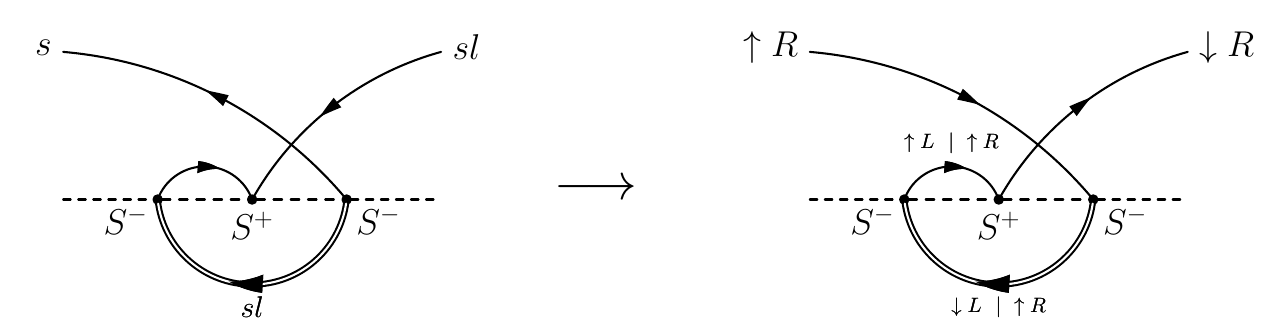}
\caption{Example of translation of a third-order, conventional-scheme, poor man's scaling diagram with no fermion loop. As always, the dashed line represents the impurity, full lines are fermions, and a double line is the scattered edge-state fermion.}
\label{fig:noloopC}
\end{figure}

Open-loop diagrams similar to those in Fig.~\ref{fig:nonloop}, by the same arguments as in the direct-model scheme, do not contribute to the flow. They continue, even in the \textit{conventional} framework, to belong to the class of \textit{reducible} diagrams, since they can be made from combining lower-order conventional-scheme diagrams. The fact that these diagrams do not contribute means that all of the possible unphysical diagrams at the third order do not affect the flow of the Kondo couplings, (since all the unphysical contributions can only come from the open-loop diagrams). 
It nevertheless might be interesting to look at an example of such a reducible unphysical diagram, which we show in Fig.~\ref{fig:noloopC}.

The diagram shown can be thought of as constructed from the first diagram in Fig.~\ref{fig:unph} plus an additional anomalous Kondo vertex, hence it is reducible. In the diagram of Fig.~\ref{fig:noloopC} we have two anomalous vertices and one normal one. Looking at the translation in terms of direct-model fermions, we can see that it asks for two propagators that involve unphysical changes (of either only channel or both spin and channel orientation) along the contraction. In addition, we can see that such a diagram would contribute to the actual rescaling of the Kondo vertex, (as opposed to the second-order unphysical diagrams that generate an additional interaction in the Hamiltonian). It is a fortunate property of the poor man's scaling procedure that these diagrams are disregarded based on the fact that they are reducible, otherwise one would have had additional unphysical contributions to the third-order flow of the coupling constant. 

\subsubsection{Wave-function renormalization}

The terms contributing to the wave-function renormalization are given by Eqs.~(\ref{eq:wfrC1}) and (\ref{eq:wfrC2}). To get the actual contribution we assume that $k=k^{\prime}$ and change from the summation in momentum to an integration in energy. The reason why the unphysical diagrams do not contribute is that, for $k=k^{\prime}$, the contributions from the terms in Eq.~(\ref{eq:unphy}) are zero. The absence of these unphysical contributions means that the wave-function renormalization is exactly the same as in the direct-scaling scheme. This can be confirmed by doing the calculation in the conventional scheme and arriving at
\begin{equation}
\begin{split}
 -\bigg(2 J^{2}_{z} \rho^{2}_{0} |\delta D|+4 J^{2}_{\perp} \rho^{2}_{0} |\delta D| \bigg ) \bigg( \ln(2) -\frac{E}{2D} \bigg )\\
\end{split}
\end{equation}
which is the same as we found in the direct-model calculation. Since the terms coming directly from the rescaling of the Kondo vertices are also the same between the two schemes, as we have seen in the previous section, this means that the final expression for the $\beta$ functions will be the same in both the conventional and direct schemes. 

Next we can proceed to do the poor man's scaling using the \textit{consistent} compactification scheme in order to compare it to the results of the direct and conventional calculations.

\subsection{Consistently compactified model}

In the \textit{conventional}-scheme calculation, products of vertex operators of the form $\exp({{i\phi_{\nu}}/{2}})\exp({{-i\phi_{\nu}}/{2}})$ are treated as standard products of exponentials with opposite-sign arguments and taken to give \textit{one} (the identity operator). 
Whereas in the \textit{consistent}-scheme calculation, we preserve those products \textit{intact} by only introducing a shorthand notation for them \cite{shah2016,*bolech2016}, 
namely, the (Fermi-density-like) operators $\tilde{n}$ defined as
\begin{equation}\label{eq:nfactors}
    \begin{split}
        & \sqrt{2}\tilde{n}_c \equiv e^{i\frac{\phi_{c}}{2}}e^{-i \frac{\phi_{c}}{2}}\\
         & \sqrt{2}\tilde{n}^{\ell}_l \equiv e^{i\ell \frac{ \phi_{l}}{2}}e^{-i\ell \frac{\phi_{l}}{2}}
    \end{split}
\end{equation}
We shall sometimes refer to these operators as \textit{consistency factors}. They satisfy the properties of idempotence, $\big ( \tilde{n}^{\pm} \big )^2 = \tilde{n}^{\pm}$, and co-nilpotence, $\tilde{n}^{+} \tilde{n}^{-}=0$, for each independent $\tilde{n}$-history \cite{Ljepoja2024a}. These properties of the $\tilde{n}$'s turn out to be exactly was is needed in order to eliminate the unphysical diagrams from the calculation of the coupling-constant flow.

Within the consistent refermionization scheme, the spin-flip part of the Hamiltonian mapped into
\begin{equation*}
\begin{split}
H_{K}^{\perp} & = \tilde{n}_{c}\tilde{n}^{-}_{l}J_{\perp} S^{-} \sum_{k,k^{\prime}}c_{k,sl}c^{\dagger}_{k^{\prime},s} \\
&\qquad - \tilde{n}_{c}\tilde{n}^{-}_{l}J_{\perp} S^{+}\sum_{k,k^{\prime}} c^{\dagger}_{k,sl}c_{k^{\prime},s}\\
&\qquad + \tilde{n}_{c}\tilde{n}^{+}_{l}J_{\perp} S^{-}\sum_{k,k^{\prime}}c_{k,sl}^{\dagger}c_{k^{\prime},s}^{\dagger}\\
&\qquad - \tilde{n}_{c}\tilde{n}^{+}_{l}J_{\perp} S^{+}\sum_{k,k^{\prime}} c_{k,sl}^{}c_{k^{\prime},s}^{}
\end{split}
\end{equation*}
whereas the parallel interaction became

\begin{equation}
\begin{split}
H^{z}_{K}& = \tilde{n}_{c}\,(\tilde{n}^{+}_{l}+\tilde{n}^{-}_{l})\,J_{z} S^{z} 
             \sum_{k, k^{\prime}} c_{k,s}^{\dagger}c_{k^{\prime},s}\\
& - \tilde{n}_{c} J_{z} S^{z}\sum_{k, k^{\prime}} 
\big( \tilde{n}^{+}_{l} c_{k^{\prime},sl}c^{\dagger}_{k,sl} 
+ \tilde{n}^{-}_{l} c_{k,sl}^{\dagger}c_{k^{\prime},sl}\big) \\
& = \tilde{n}_{c} \tilde{n}^{-}_{l} J_{z} S^{z} \sum_{k, k^{\prime}}\big ( c_{k,s}^{\dagger}c_{k^{\prime},s}-c_{k,sl}^{\dagger}c_{k^{\prime},sl} \big )\\
& + \tilde{n}_{c} \tilde{n}^{+}_{l}J_{z} S^{z}\sum_{k, k^{\prime}} \left( c_{k,s}^{\dagger}c_{k^{\prime},s} - c_{k^{\prime},sl}c^{\dagger}_{k,sl} \right)
\end{split}
\end{equation}

In this \textit{consistent} framework, anomalous and normal vertices of the spin-flip part of the interaction are separated in such a way that normal vertices come multiplied with $\tilde{n}^{-}_{l}$, while  the anomalous ones come with $\tilde{n}^{+}_{l}$. This fact prevents the mixing of anomalous and normal vertices because of the co-nilpotence of the consistency factors. That means that there will not be any diagrams like the ones in group (B) of Fig.~\ref{fig:2ndPMC}. This is one of the main outcomes of preserving of the consistent consistency factors: they prevent otherwise allowed unphysical diagrams from contributing to the calculations. Additionally, in comparison with the conventional scheme, one notices that the $J_{z}$ vertex scatters both $s$- and $sl$-sector fermions (mirroring the parallel interaction in the direct scheme). 

\subsubsection{Second-order scaling}

For the second-order contribution we can draw the same diagrams as the ones shown in Fig.~\ref{fig:2ndPMC}, but excluding the unphysical ones in group (B), since the consistency factors make their contributions vanish. The diagrams in groups (C) and (D) are still contributing to the flow of $J_{z}$ and $J_{\perp}$, respectively, while the diagrams in group (E) still contribute to the flow via the wave-function renormalization. 

In addition, we also have the diagrams in group (A) contributing to the flow of $J_z$. Those diagrams, in the conventional-scheme calculation, were canceling between the two rows. However, in the consistent-scheme case, the diagrams in the first (second) row come multiplied by $\tilde{n}^{-}$ ($\tilde{n}^{+}$) and this difference prevents the cancellation between them. These diagrams are going to contribute to the flow of $J_z$ (and help preserve the \textit{consistent} $s$-$sl$ balance of the parallel Kondo interaction). The calculation of these diagrams (using fermionic contractions and spin identities), proceeds in the same way as in the conventional- and direct-scheme calculations.

It is important, however, to illustrate at this point how the $\tilde{n}$'s are treated in the consistent-scheme. For that, we can first take a look at the diagrams in group (C) as an example. The diagrams in the top row of this group are ``anomalous'', (meaning that they combine two anomalous vertices). Since each anomalous vertex comes multiplied by $\tilde{n}_{l}^{+}$, the contribution from these diagrams will be multiplied by $ \tilde{n}_{l}^{+} \tilde{n}_{l}^{+}$. On the other hand, the diagrams in the bottom row are ``normal'' and their contribution is multiplied by $\tilde{n}_{l}^{-} \tilde{n}_{l}^{-} $. Other than those factors, the diagrams coming from contracting two \textit{normal} or two \textit{anomalous} vertices contribute equally. Calling upon the idempotence property  ($\tilde{n}^2=\tilde{n}$) of the adiabatically conserved local fermionic densities, we have that the contribution from these diagrams is given by
\begin{equation}
\begin{split}
& 2 J^{2}_{\perp}\rho_{0} \frac{|\delta D|}{D} \tilde{n}_{c} \tilde{n}^{-}_{l} S^{z}\sum_{k, k^{\prime}} c_{k,s}^{\dagger}c_{k^{\prime},s}\\
& \qquad + 2 J^{2}_{\perp}\rho_{0} \frac{|\delta D|}{D} \tilde{n}_{c} \tilde{n}^{+}_{l} S^{z}\sum_{k, k^{\prime}} c_{k,s}^{\dagger}c_{k^{\prime},s}
\end{split}
\end{equation}
This gives the flow of $J_{z}$ for the fermion-impurity parallel scattering in the $s$-sector. In order to check if the $sl$-sector part consistently flows in the same way, we need to include also the diagrams in group (A) of Fig.~\ref{fig:2ndPMC}. As opposed to the conventional-scheme case, --where those diagrams were mutually canceling each other between the two rows--, in the consistent-scheme case the two rows come with different $\tilde{n}$ factors, and so the cancellation is avoided. As a result, they separately, but in the same manner, contribute to the scaling of $J_{z}$  and the two parts of the vertex flow as one (meaning they have the same flow). Therefore, the full expression of the scaling of the $J_{z}$ coupling constant, coming from both group-(A) and group-(C) diagrams is given by
\begin{equation}
\begin{split}
& 2 J^{2}_{\perp}\rho_{0} \frac{|\delta D|}{D} \tilde{n}_{c} \tilde{n}^{-}_{l} S^{z} \sum_{k, k^{\prime}}\big ( c_{k,s}^{\dagger}c_{k^{\prime},s}-c_{k,sl}^{\dagger}c_{k^{\prime},sl} \big )\\
& \quad + 2 J^{2}_{\perp}\rho_{0} \frac{|\delta D|}{D} \tilde{n}_{c} \tilde{n}^{+}_{l} S^{z}\sum_{k, k^{\prime}} \big ( c_{k,s}^{\dagger}c_{k^{\prime},s} - c_{k^{\prime},sl}c^{\dagger}_{k,sl} \big )\\
\end{split}
\end{equation}
This preserves the $s$-$sl$ structure of the vertex and gives us the second-order flow of the parallel coupling constant, $J_{z}$, to be the same as the one we obtained from both of the two previous calculations. The same outcome is found when calculating the second-order flow of the $J_{\perp}$ coupling constant. It is important to remark that in the consistent scheme, since mixing of the anomalous and normal vertices is prohibited, there can be no additional unphysical interactions appearing in the model like the ones in Eq.~(\ref{eq:unphy}). This is an advantage of the consistent scheme as opposed to the conventional one.

\subsubsection{Third-order flow}

The third-order diagrams contributing to the flow of the coupling constants in the consistently compactified model are the same as those already given in Fig.~\ref{fig:3rdPMC} for the conventional case. Even the diagrams in group (B), that seemingly combine normal and anomalous vertices, are valid contributions; the reason being that the vertices forming the fermion loop can have either $\tilde{n}^{-}$ or $\tilde{n}^{+}$ at them regardless of the consistency factor assigned to the middle vertex of the diagram. This is because, when doing the fermionic loop contraction, one does the trace over its vertices before multiplying by the remaining $\tilde{n}$ factors and utilizing their co-nilpotence property. In addition, when it comes to the actual contribution coming from the fermionic contractions in the third-order diagrams, they are independent of the $\tilde{n}$'s at the outer diagram vertices. To give an illustration of how the calculation proceeds let us call that contribution (modulo consistency factors) $\Gamma_{3}$ and take the example of the middle vertex being a normal one (it would be the same for an anomalous one, just with $\tilde{n}_{l}^{+}$ instead as the common central-vertex factor). Diagrams with normal and anomalous outer vertices will give the following combined contribution
\begin{equation}\label{eq:cons3rd}
    \begin{split}
    & \mathrm{Tr} \ \big ( \tilde{n}_{l}^{-} \cdot \tilde{n}_{l}^{-} \big )\ \tilde{n}_{l}^{-} \times \Gamma_{3}+\mathrm{Tr} \ \big ( \tilde{n}_{l}^{+} \cdot \tilde{n}_{l}^{+} \big ) \ \tilde{n}_{l}^{-} \times \Gamma_{3}\\
    & \qquad = \, \mathrm{Tr} \ \big ( \tilde{n}_{l}^{-} + \ \tilde{n}_{l}^{+} \big ) \ \tilde{n}_{l}^{-} \times \Gamma_{3}
    \, = \, 2 \ \tilde{n}_{l}^{-} \times \Gamma_{3}
    \end{split}
\end{equation}
In obtaining this result, we used the properties of the consistency factors that $\tilde{n}^{\pm} \tilde{n}^{\pm} = \tilde{n}^{\pm}$ and also $\tilde{n}^{+} + \ \tilde{n}^{-} = 1$. The mechanism of how the consistent scheme allows for the mixing of normal and anomalous vertices in the third-order fermion-loop diagrams is now evident. As one can see from the above equation, the vertices involved in the fermion loop are under a trace operator, and that prevents the direct multiplication of their consistency factor(s) by additional $\tilde{n}$ factors which would seem, in certain cases, to produce a zero based on the co-nilpotence property, [the case of the second term in Eq.~(\ref{eq:cons3rd})]. The fact that one can mix vertices like that in these third-order diagrams has an elegant interpretation if one thinks of the comparison to the direct-model calculation. In that case, we had the $L$ and $R$ leads, and, since there is no inter\-lead scattering, those labels are not mixed in any of the above-considered diagrams except for the third-order fermionic-loop diagrams (or any diagrams at higher orders including fermionic loops in general) where, in a loop, one can have either $L$ or $R$ fermions regardless of the lead assignment of the external fermion legs. To put it differently, fermionic loops count the number of leads (which is just $2$ in the two-lead case), and that is what the trace of the identity gives in the consistent framework [the last step in Eq.~(\ref{eq:cons3rd})]. Indeed, one can back-translate the $\tilde{n}_{l}^{-}$ terms in the Hamiltonian as corresponding to the $L$ lead, and the $\tilde{n}_{l}^{+}$ ones to the $R$ lead. Then the fact that one can have both normal and anomalous vertices in a loop is needed to count the number of ``channels'' in the model.

Calculating all of the third-order diagrams, in the way described above, we arrive at exactly the same contribution as we had with the direct formulation of the model. The flow of the Kondo coupling constants will thus be the same between the two schemes (as expected, since the consistent compactification does not introduce unphysical processes, nor does it under- or over-count the physical ones).

\section{Universality}

The result of comparing the direct calculation and the two different bosonization-debosonization schemes is that the $\beta$ function(s) for the Kondo coupling constant(s) comes out to be the same in all three cases, namely,
\begin{equation}
\begin{split}
\beta(g) = -g^2 + M g^{3}
\end{split}
\end{equation}
where we have quoted the result for the fully isotropic case to keep the discussion simpler. (The conclusions would be similar in the case with spin anisotropy, since it turns out to be an \textit{irrelevant perturbation} in the Wilsonian sense.) Recall that we are calling $M$ the number of channel pairs and one can perform a pairwise compactification of the original Kondo model. In the large-$M$ limit, the infrared fixed point is perturbatively accessible as the (first) zero of the flow toward strong coupling, and it is reached for $g^{\star} = 1/M$. 

\begin{figure*}[t]
\includegraphics[width=0.9\textwidth]{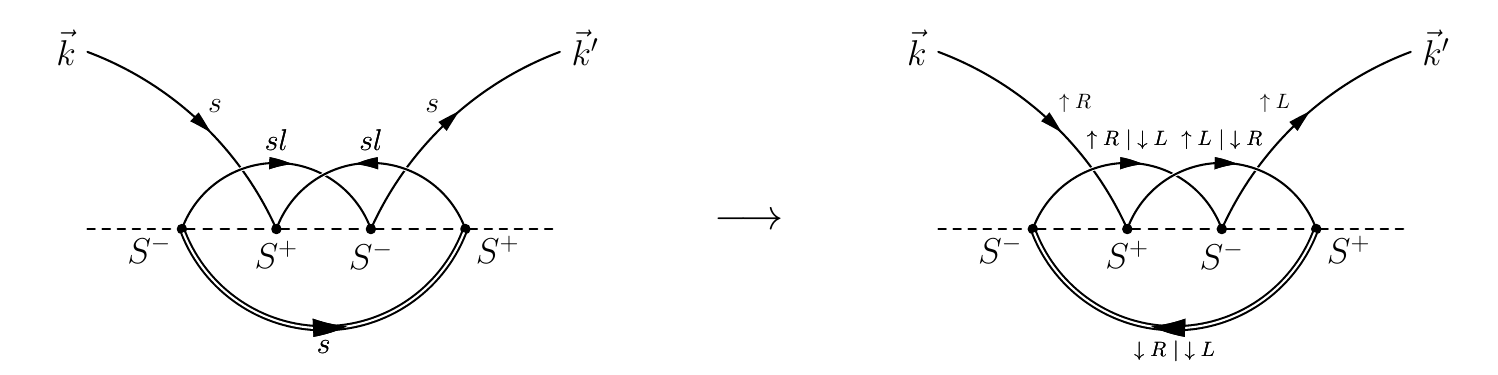}
\caption{Example of translation attempt of a fourth-order, conventional, poor man's scaling diagram with no fermion loop. As always, the dashed line represents the impurity, full lines are fermions and a double line is the scattered edge-state fermion (joining the two outer vertices). One can see that this particular example would translate into an unphysical process in the direct language. In other words, this conventional diagram is \textit{untranslatable}. Furthermore, one would be unable to make a diagram with such a topology/connectivity and all spin-flip vertices in the direct language. It would require connecting two $S^{+}$ or two $S^{-}$ operators with a fermion propagator, which in the original formulation of the model is not allowed (for spin-1/2 band fermions).}
\label{fig:4th}
\end{figure*}

These coincident results for the beta function are as expected; since we have seen that there are no unphysical (compactified-scheme) diagrams involved in the scaling of the Kondo interaction up to the third order in the coupling constant. In contrast, going to one-higher order one recognizes that there will be two distinct sets of contributions. One of them will comprise the contributions from diagrams with fermionic loops, (\textit{i.e.}, $M$-dependent contributions). These have been calculated in the literature \cite{gan1993,gan1994,fischer1999}, --but using field-theoretical methods that we shall discuss later \cite{Ljepoja2024c}. They do not involve untranslatable contributions, since the presence of fermionic loops still severely restricts the unphysical combination of anomalous and normal vertices (at the leading order in $M$ if one moves to even-higher orders in the coupling constant) and are delicate to access within a T-matrix-based approach to scaling. The other set of contributions comes from diagrams of order $J^{4}$ without any fermionic loops, and thus no $M$ dependence; cf.~Ref.~\onlinecite{affleck2005}. An (untranslatable) example of such a diagram is shown in Fig.~\ref{fig:4th}, together with its attempted translation into direct-model language
\footnote{It is important to notice that diagrams of the type of that in Fig.~\ref{fig:4th} are the only ones that contribute to the channel-independent part of the scaling, since they are the only irreducible fourth-order diagrams without fermionic loops. However, besides the $J_{\perp}^4$ unphysical contribution to the flow of $J_{z}$, there is also a $J_{\perp}^3 J_{z}$ contribution to the flow of $J_{\perp}$ which is also unphysical and untranslatable; and those are the only two untranslatable contributions at this order of perturbation (all the others are physical instances of this diagram that have a correlate in the direct calculation of the beta function; cf.~Appendix B of Ref.~\onlinecite{affleck1991}).}. 
One can see that such a diagram would contribute to the rescaling of the parallel coupling constant in the conventional scheme, while in the direct-scheme language it would correspond to a diagram with (in this case both) the channel flavor or the spin orientation being flipped along a single fermion propagator (which is unphysical, as we already argued at length above).

Even though they are unphysical, diagrams like the one in Fig.~\ref{fig:4th} will alter the beta function found in the calculation based on the conventionally compactified model. One would like to know if and how that change affects the low-temperature universal aspects of the physics; (by modifying, for instance, the thermodynamic scaling exponents). To that end, let us assume that we know the fourth-order (in $g$) contribution, and the resulting flow is given by
\begin{equation}\label{eq:flow}
    \begin{split}
        \beta(g) = -g^2 + M g^{3}+a g^{4}
    \end{split}
\end{equation}
where the fourth-order coefficient $a$ takes different numerical values according to the conventional- or the direct/consistent-scheme calculations (since, as we have just seen, the conventional scheme includes unphysical contributions to that coefficient
\footnote{The beta-function coefficient $a$ will have two parts: a channel-dependent one, which will be the same in all schemes and given by $(1+\mathrm{\ln(2)})M$, plus a channel-independent part that will differ between the schemes, (the direct-scheme calculation will give $\pi^2 / 4$, whereas the alternative value will be $11 \pi^2 / 64$ for the conventional-scheme).}). 

One can then find the infrared fixed point, $g^{\star}$, by solving for $\beta(g^{\star}) = 0$. One way of determining the shifted fixed point location is by applying Newton's root-finding method, where for each/(the first) iteration we use the previous/(third-order) fixed-point location, ($1/M$), and we would iterate for each added order in the expansion of the beta function. Following that procedure we arrive at $g^{\star} = (1 - a_{1})/M + O(a^{2}_{1})$, where $a_{1} = a/M^2$. Here we have assumed that $a_{1}$ is small and expanded to keep only the linear terms. Notice, however, that we did not assume that the channel number is particularly large, we assumed only that $M > 1$ in order to study the behavior around a perturbatively accessible intermediate-coupling fixed point (the expansion will, of course, be more accurate the larger $M$ is). The derivative of the beta function at the fixed point is then $\Delta \equiv \beta'(g^{\star}) = 1/M + O(a^2_{1})$. As there is no linear correction, this is an unchanged result from the previous/third-order, and we shall see that this fact is responsible for the relative insensitivity of the (``universal'') thermodynamic scaling exponents \cite{Barzykin1998}.

The first thing we need to do now, in order to study the behavior near the infrared fixed point, is to define the (crossover) Kondo temperature, $T_{K}$, and figure out the running of the coupling constant. These can be done by separating variables in the definition of the beta function and integrating:
\begin{equation}
    \begin{split}
        \int_{D}^{\omega} d (\ln (\omega^{\prime})) = \int_{g_{0}}^{g_{r}} \frac{d g^{\prime}}{\beta (g^{\prime})}
    \end{split}
\end{equation}
where $D$ is the energy cutoff or band edge, $g_{r} (\omega)$ is the running coupling constant at energy $\omega$, and $g_{0}$ is the bare coupling constant. The integral on the r.h.s.~of the equation is not as easy to compute as in the case of the third-order beta function. But since we are interested in the behavior of the model around the fixed point ($g^{\star}$), we can make some approximations that are accurate near the relevant fixed points (notice that further-removed fixed points, appearing due to the higher orders in the expansion, are unphysical in any case since the flow can never reach them starting from the Gaussian fixed point). Namely, we can approximate the beta function as
\begin{equation}\label{eq:flow}
    \begin{split}
        \beta(g) \approx \Delta \bigg( \frac{g}{g^{\star}} \bigg)^2 (g-g^{\star})
    \end{split}
\end{equation}
Basically, we replace the original higher-order-in-$g$ corrections to the flow with a lower-order interpolant that has the same location-of and slope-at $g^{\star}$ as the higher-order beta function does, (\textit{i.e.}, $g^{\star}$ and $\Delta$, respectively). Of course, it also has the second-order zero at $g=0$ that corresponds to the ultraviolet fixed point. The integral is now again elementary and the equation for the running coupling becomes
\begin{equation}
    \begin{split}
        \ln \bigg( \frac{\omega}{D} \bigg ) & = \frac{ ( g^{\star} )^2}{\Delta}\int_{g_{0}}^{g_{r}} \frac{d g^{\prime}}{(g^{\prime})^2 (g^{\prime}-g^{\star})}\\
        & = \frac{g^{\star}}{g_{r}\Delta} - \frac{g^{\star}}{g_{0}\Delta} + \frac{1}{\Delta} \ln \bigg [\frac{(g^{\star}-g_{r}) g_{0}}{(g^{\star}-g_{0})g_{r}  } \bigg ]
    \end{split}
\end{equation}
From this equation we can find the Kondo temperature by using that it can be (conventionally) defined as the temperature at which the running coupling constant becomes $\frac{2}{3}g^{\star}$. This corresponds to the local minimum of the beta function that lies in between the UV and IR fixed points
\footnote{One could also more simply adopt $g^\star/2$, the mid point between the two fixed points. Such an alternative convention would result in the same leading exponential behavior, $\exp(-g^\star/g_0\Delta)$, and slightly simpler numerical factors, but is less well physically motivated.}.
Using this conventional definition for the Kondo temperature, and units in which the Boltzmann constant was set to one, we have
\begin{equation}
    \begin{split}
        \ln \left( \frac{T_{K}}{D} \right)& = \frac{3/2}{\Delta} - \frac{g^{\star}}{ g_{0}\Delta} + \ln \bigg [\frac{g_{0}/2}{g^{\star}-g_{0}} \bigg ]^{\frac{1}{\Delta}}
    \end{split}
\end{equation}
Choosing $g_{0}\ll g^{\star}$, the equation simplifies and we can solve for the Kondo scale
\begin{equation}
\begin{split}
T_{K} & = D e^{-\frac{g^{\star}/\Delta}{g_{0}}+\frac{3}{2\Delta}} \left( \frac{g_{0}/2}{g^{\star}-g_{0}} \right)^{\frac{1}{\Delta}}
\end{split}
\end{equation}
Now one can use this expression as a boundary condition to write the running coupling constant for $\omega\ll T_{K}$ as
\begin{equation}\label{eq:run.coupl}
\begin{split}
g_{r}(\omega) &= g^{\star} - \frac{1}{2} g_{r}(\omega) e^{\frac{3}{2}-\frac{g^{\star}}{g_{r}(\omega)}} 
\left(\frac{\omega}{T_{K}}\right)^{\Delta} \\
&\simeq \frac{g^{\star}}{1+\frac{\sqrt{e}}{2}
\left({\omega}/{T_{K}}\right)^{\Delta}}
\end{split}
\end{equation}

This is a very generic type of behavior also found in other asymptotically free theories \cite{Appelquist1998}.
We can see that the frequency scaling exponent is given by $\Delta$, which at linear order is not affected by the fourth-order contributions to the beta function. This means that the exponent is going to be the same (at this level of approximation) whether we use the consistent or the conventional schemes, even though the conventional scheme does have unphysical terms contributing. Therefore, as far as the leading-order universality of the low-temperature physics is concerned, conventional and consistent schemes would still produce the same generic results, even though their beta functions start to differ. 

Indeed, now that we have the running coupling constant, we can proceed further and use it to calculate the impurity entropy and specific heat. To do so we start from the perturbatively obtained impurity contribution to the free energy that is given by
\small
\begin{equation}
    F_\mathrm{imp} (T) = -E_{0} -T \ln(2) +\frac{1}{2} \pi^2 T ( M g^3 -\frac{3}{4} M^{2} g^4) + O(g^4)
\end{equation}
\normalsize
This result was obtained in the literature using a diagrammatic expansion and keeping only terms that are linear in temperature \cite{Kondo1968,gan1994}. Due to the scaling invariance of the impurity free energy, we can replace the $g$ appearing in its expression with the running coupling constant corresponding to some frequency scale given by the temperature $T$, namely, $g_{r}(T)$. The temperature that we choose is in the range of the Kondo temperature but still smaller than it, $T\lesssim T_{K}$. This enables us to keep only the lower-order terms in the temperature expansion. (This assumption is important, as we shall see, in determining which orders are kept and which suppressed.) The working expression for the free energy thus becomes
\begin{widetext}
\begin{equation}
\begin{split}
F_\mathrm{imp} (T) & = -E_{0}-T \ln(2) +\frac{1}{4} \pi^2 T \bigg( 2 M g^{3}_{r}(T) - \frac{3}{2} M^2 g^4_{r} (T)  \bigg ) + O(g^{4}_{r} (T))\\
& \approx -E_{0}-T \ln(2) +\frac{1}{4} \pi^2 T \bigg [ \frac{1}{2 M^2} - 6\frac{a_{1}}{M} \xi\bigg( \frac{T}{T_{K}} \bigg)^{\Delta} - \big ( 3- 12 a_{1} \big ) \xi^2 \bigg ( \frac{T}{T_{K}}\bigg )^{2 \Delta}  \bigg ] + O(T (T/T_{K})^{3 \Delta})\\
\end{split}
\end{equation}
\end{widetext}
where we used the expression for $g_{r}(T)$ taken from Eq.~(\ref{eq:run.coupl}) and introduced the constant $\xi = \sqrt{e} g^{\star} / 2$. In addition, we replaced $g^{\star} = (1-a_{1})/M$ in the final expression. Inspecting the resulting free energy, one identifies three terms that are important to consider. The first one is the term that is linear in $T$. This term contributes to the residual impurity entropy, $S_0=\ln(2)-\pi^2/8M^2$, and does not have any $a_1$ dependence. The other two are the terms that are linear and quadratic in $(T/T_{K})^{\Delta}$. Since the temperature is in a small range just below $T_{K}$, both of these terms are comparable in magnitude. However, the term that is linear in $(T/T_{K})^{\Delta}$ is multiplied by ${a_{1}}/{M}$, which is a small coefficient even for a finite $M>1$. Additionally, since $M \sim 1/ \sqrt{a_{1}}$, we know that $a_{1}/M \sim a^{3/2}_{1}$, which is beyond the linear-order approximation in $a_{1}$ used so far. Therefore, the term is suppressed and should be dropped (unless one introduces even higher powers of the coupling in the analysis, since those can be seen to contribute at a similar order in $a_{1}$ to the next-order correction of the slope of the beta function at the fixed point).

Having determined the leading contribution to the impurity free energy, one can then calculate the impurity specific heat to find
\begin{equation}
    \begin{split}
       & C_\mathrm{imp} \approx \frac{3-12 a_{1}}{2} \pi^2 \xi^2 \Delta (1+2\Delta)\bigg ( \frac{T}{T_{K}} \bigg )^{2 \Delta}\\
    \end{split}
\end{equation}
Inspecting this expression, one can see that $a_{1}$ enters only in the multiplicative coefficient, but not in the scaling exponent which stays the same, $\alpha = 2 \Delta$, as it would have been without including the fourth-order contribution to the beta function. In addition, the result is in fair agreement with that obtained perturbatively in the large-$M$ limit \cite{gan1993,gan1994,bensimon2006}, as well as with Bethe ansatz \cite{andrei1984,bolech2002,*bolech2005a} and Conformal Field Theory \cite{affleck1991,johannesson2003,*johannesson2005} calculations. This further solidifies the picture that the unphysical diagrams in the conventional scheme, like the one in Fig.~\ref{fig:4th}, do contribute to the flow at higher orders, but not to the universal aspects of the physics around the IR fixed point of the model.

Another interesting situation in which we can compare the two compactification schemes is with the addition of channel anisotropy. This is better visualized in the context of a quantum-dot realization of the model \cite{oreg2003,*potok2007,bolech2005b,shah2006,mitra2011}. The electrons in our formulation of a ``mesoscopic'' multichannel Kondo model will then have a total of $2M$ \textit{microscopic} channels or flavors (in addition to the spin). We will have two \textit{macroscopic} groups of flavors, that we can refer to as the \textit{leads} ($L$ and $R$) of a \textit{gedanken} transport-experiment setup, and they each will have a further $M$-fold degeneracy (the ``conduction channels'' of each lead in the mesoscopic setting), giving a total of $2M$ flavors. It is on this lead-flavor index ($L$ or $R$) that the coupling constant will depend on in the channel-anisotropic case that we want to consider here 
\footnote{For simplicity, we will not add a co-tunneling term. In that way the model reduces, at the channel-symmetric point, to the one we have been studying so far, and, moreover, the setting is similar to that of the so-called ``charge Kondo'' model \cite{matveev1991,*matveev1995,*Furusaki1995,Iftikhar2015,*Iftikhar2018,Karki2022,Pouse2023}, which was used to experimentally demonstrate the RG flow with channel asymmetry.}.
Thus, for each of the $M$-degenerate conduction channels, we have the $g^{L}$ and $g^{R}$ coupling constants that can be different from each other. The reason why we choose this way of defining a multichannel model is in order to obtain a simpler (compact) model in terms of the physical flavors ($c$, $l$, $s$, and $sl$) introduced after the bosonziation procedure. Carrying out all the calculation in the direct scheme, we arrive at the spin- and channel-anisotropic flow given by the beta functions
\begin{equation}
    \begin{split}
        & \beta^{\alpha}_{\perp} = \ - \ g^{\alpha}_{z}g^{\alpha}_{\perp} + \frac{M}{4} g^{\alpha}_{\perp}  \big( g^{\alpha}_{z} \big)^2  + \frac{M}{4} g^{\alpha}_{\perp}\big ( g^{\bar{\alpha}}_{z} \big)^2\\
        & \qquad \qquad + \frac{M}{4}  g^{\alpha}_{\perp}\big( g^{\bar{\alpha}}_{\perp}\big)^2 + \frac{M}{4}  \big(g^{\alpha}_{\perp}\big)^3\\
        & \beta^{\alpha}_{z} = \  -\big( g^{\alpha}_{\perp} \big)^2 + \frac{M}{2} g^{\alpha}_{z} \big( g^{\alpha}_{\perp} \big)^2 + \frac{M}{2} g^{\alpha}_{z}\big( g^{\bar{\alpha}}_{\perp} \big)^2\\
    \end{split}
\end{equation}
where $\bar{\alpha}$ takes the values $L$ or $R$ when $\alpha$ is $R$ or $L$, respectively. This is the result for the most general case of a (multi) two-channel model with spin and channel anisotropy. We checked that it correctly recovers previous poor man's scaling calculations that have been carried out only for less general cases and/or to lower orders \cite{Anderson1970,Hewson,Nevidomskyy2015,kuramoto1998,Solyom1974}.

Performing the calculation in the conventional scheme, one will recover the exact same result. For the multichannel model, defined as we have, one can do the rotation into the physical sectors for each of the $M$ conduction channels. In the conventional scheme $g^{L}$ describes the coupling of the normal vertex with the impurity, while $g^{R}$ goes with the anomalous coupling. Mixing of these two types of processes can lead to unphysical contributions, but not until higher orders are reached. 

\section{Interim Conclusion and Summary}

In concluding, let us first recapitulate our findings regarding the bosonization-debosonization (BdB) procedure that was used to compactify the (multi) two-channel Kondo model. On the one hand, the comparative results of poor man's scaling confirm our earlier findings for the consistent BdB of the model \cite{Ljepoja2024a}; namely, that the \textit{consistently} compactified model is an exact reparameterization of the original multichannel Kondo Hamiltonian in terms of an alternative set of fermions, each separately associated with the model's different \textit{physical sectors}. Along the way, our next-to-leading-order diagrammatic calculations helped clarify the role and properties of the \textit{consistency factors} given by the $n$-twiddle operators which are crucial to enable the compact reparameterization to be exact. 

On the other hand, the \textit{conventionally} compactified model, quite remarkably, gives the same results for the beta function(s) up to the order computed here (which, some could argue, captures all the universal aspects of the low-energy physics) and deviates only at higher orders. Correspondingly, the low-energy thermodynamics near the infrared fixed point of the model is also identical in all the calculational schemes considered. This is in contradistinction with our earlier findings that certain exact results involving finite systems or special limits are different for the conventional scheme (in particular those involving nonequilibrium transport calculations) \cite{Ljepoja2024a}. The reason for the surprising agreement of the beta functions is that the dynamics of the impurity itself restricts the appearance (or manifestation) of unphysical processes until higher orders of perturbation are reached. This serendipity implies that for other types of impurities one would need to check, on a case-by-case basis, if similar restrictions are also present or not before trusting results involving the \textit{conventional} use of BdB in the calculations. Moving beyond Abelian bosonization, it would be natural to ask, in further studies, how and to what extent are these findings mirrored in the non-Abelian formalism.

The caveats of the previous paragraph notwithstanding, the conventional way of compactifing the (multi) two-channel Kondo model still constitutes an appealing approach to the study and possible realization of (local) non-Fermi liquid physics in impurity-based systems. Indeed, the original/conventional compactification was not claimed to be exact. Rather, it was supposed to capture only the most desirable aspects of the low-energy physics near the IR fixed point \cite{coleman1995b}. Our findings are in agreement with that scenario provided the systems are probed in near-equilibrium conditions. It had also been argued that compactification moves the IR fixed point of the two-channel Kondo impurity ($M\!=\!1$) from intermediate to strong coupling and turns channel asymmetry into a marginal operator (thus turning the limiting separatrix that flows from the two- to the one-channel fixed points into a strong-coupling fixed line \cite{coleman1995}). Since our scaling calculations are perturbative, we cannot confirm that scenario. However, for $M\!>\!1$, both the $L$-$R$ symmetric and asymmetric cases have intermediate-coupling fixed points that are perturbatively accessible. We find in these cases that the \textit{conventional} compactification of the model does not change the beta function even in the presence of channel asymmetry, which remains a relevant operator.

We thus saw that the study of the compactification of the (multi) two-channel Kondo model was valuable to further our understanding of the BdB formalism. In that framework, the comparative study of scaling allowed us to identify the limits of validity of the \textit{conventional} implementations of BdB-based mappings. Interestingly, those limits turned out to coincide with the conventional wisdom for the limits of universality of Kondo models. However, so far we argued only qualitatively about those limitations. In order to do it quantitatively, we would need to go further and study the next-to-next leading orders of scaling, \textit{i.e.}, the leading nonuniversal corrections. That is a task that falls outside the practical use of the poor man's scaling approach presented here. Instead, one needs to resort to a more systematic implementation of the RG program using field-theoretic methods (but such a calculation is more technically involved and will require a separate presentation \cite{Ljepoja2024c}). Let us anticipate that, to the same order for the beta functions as considered here, we shall see that the field-theoretic RG calculations exactly validate our current results.

\acknowledgements  
We are grateful for the hospitality of Pune's Indian Institute for Science Education and Research (IISER-Pune) and the Harvard-Smithsonian Institute for Theoretical Atomic Molecular and Optical Physics (ITAMP), which hosted some of us while part of this work was taking place. We also acknowledge regular day visits and office space at the Physics departments of Harvard and Northeastern University during the same period.

\%bibliography{books2022,kondo2022,hubbard2022,tunneling2022}


\begin{thebibliography}{72}%
\makeatletter
\providecommand \@ifxundefined [1]{%
 \@ifx{#1\undefined}
}%
\providecommand \@ifnum [1]{%
 \ifnum #1\expandafter \@firstoftwo
 \else \expandafter \@secondoftwo
 \fi
}%
\providecommand \@ifx [1]{%
 \ifx #1\expandafter \@firstoftwo
 \else \expandafter \@secondoftwo
 \fi
}%
\providecommand \natexlab [1]{#1}%
\providecommand \enquote  [1]{``#1''}%
\providecommand \bibnamefont  [1]{#1}%
\providecommand \bibfnamefont [1]{#1}%
\providecommand \citenamefont [1]{#1}%
\providecommand \href@noop [0]{\@secondoftwo}%
\providecommand \href [0]{\begingroup \@sanitize@url \@href}%
\providecommand \@href[1]{\@@startlink{#1}\@@href}%
\providecommand \@@href[1]{\endgroup#1\@@endlink}%
\providecommand \@sanitize@url [0]{\catcode `\\12\catcode `\$12\catcode `\&12\catcode `\#12\catcode `\^12\catcode `\_12\catcode `\%12\relax}%
\providecommand \@@startlink[1]{}%
\providecommand \@@endlink[0]{}%
\providecommand \url  [0]{\begingroup\@sanitize@url \@url }%
\providecommand \@url [1]{\endgroup\@href {#1}{\urlprefix }}%
\providecommand \urlprefix  [0]{URL }%
\providecommand \Eprint [0]{\href }%
\providecommand \doibase [0]{https://doi.org/}%
\providecommand \selectlanguage [0]{\@gobble}%
\providecommand \bibinfo  [0]{\@secondoftwo}%
\providecommand \bibfield  [0]{\@secondoftwo}%
\providecommand \translation [1]{[#1]}%
\providecommand \BibitemOpen [0]{}%
\providecommand \bibitemStop [0]{}%
\providecommand \bibitemNoStop [0]{.\EOS\space}%
\providecommand \EOS [0]{\spacefactor3000\relax}%
\providecommand \BibitemShut  [1]{\csname bibitem#1\endcsname}%
\let\auto@bib@innerbib\@empty
\bibitem [{\citenamefont {Coleman}\ \emph {et~al.}(1995)\citenamefont {Coleman}, \citenamefont {Ioffe},\ and\ \citenamefont {Tsvelik}}]{coleman1995b}%
  \BibitemOpen
  \bibfield  {author} {\bibinfo {author} {\bibfnamefont {P.}~\bibnamefont {Coleman}}, \bibinfo {author} {\bibfnamefont {L.~B.}\ \bibnamefont {Ioffe}},\ and\ \bibinfo {author} {\bibfnamefont {A.~M.}\ \bibnamefont {Tsvelik}},\ }\bibfield  {title} {\bibinfo {title} {Simple formulation of the two-channel {K}ondo model},\ }\href {https://doi.org/10.1103/PhysRevB.52.6611} {\bibfield  {journal} {\bibinfo  {journal} {Phys. Rev. B}\ }\textbf {\bibinfo {volume} {52}},\ \bibinfo {pages} {6611} (\bibinfo {year} {1995})}\BibitemShut {NoStop}%
\bibitem [{\citenamefont {Zhang}\ and\ \citenamefont {Hewson}(1996{\natexlab{a}})}]{Zhang1996}%
  \BibitemOpen
  \bibfield  {author} {\bibinfo {author} {\bibfnamefont {G.-M.}\ \bibnamefont {Zhang}}\ and\ \bibinfo {author} {\bibfnamefont {A.~C.}\ \bibnamefont {Hewson}},\ }\bibfield  {title} {\bibinfo {title} {Linear temperature dependence of electrical resistivity in a single-impurity model},\ }\href {https://doi.org/10.1103/PhysRevLett.76.2137} {\bibfield  {journal} {\bibinfo  {journal} {Phys. Rev. Lett.}\ }\textbf {\bibinfo {volume} {76}},\ \bibinfo {pages} {2137} (\bibinfo {year} {1996}{\natexlab{a}})}\BibitemShut {NoStop}%
\bibitem [{\citenamefont {Zhang}\ and\ \citenamefont {Hewson}(1996{\natexlab{b}})}]{Zhang1996a}%
  \BibitemOpen
  \bibfield  {author} {\bibinfo {author} {\bibfnamefont {G.-M.}\ \bibnamefont {Zhang}}\ and\ \bibinfo {author} {\bibfnamefont {A.~C.}\ \bibnamefont {Hewson}},\ }\bibfield  {title} {\bibinfo {title} {Non-{F}ermi-liquid theory of a compactified {A}nderson single-impurity model},\ }\href {https://doi.org/10.1103/PhysRevB.54.1169} {\bibfield  {journal} {\bibinfo  {journal} {Phys. Rev. B}\ }\textbf {\bibinfo {volume} {54}},\ \bibinfo {pages} {1169} (\bibinfo {year} {1996}{\natexlab{b}})}\BibitemShut {NoStop}%
\bibitem [{\citenamefont {Yang}(1989)}]{Yang1989}%
  \BibitemOpen
  \bibfield  {author} {\bibinfo {author} {\bibfnamefont {C.~N.}\ \bibnamefont {Yang}},\ }\bibfield  {title} {\bibinfo {title} {\ensuremath{\eta} pairing and off-diagonal long-range order in a {H}ubbard model},\ }\href {https://doi.org/10.1103/PhysRevLett.63.2144} {\bibfield  {journal} {\bibinfo  {journal} {Phys. Rev. Lett.}\ }\textbf {\bibinfo {volume} {63}},\ \bibinfo {pages} {2144} (\bibinfo {year} {1989})}\BibitemShut {NoStop}%
\bibitem [{\citenamefont {Maldacena}\ and\ \citenamefont {Ludwig}(1997)}]{maldacena1997}%
  \BibitemOpen
  \bibfield  {author} {\bibinfo {author} {\bibfnamefont {J.~M.}\ \bibnamefont {Maldacena}}\ and\ \bibinfo {author} {\bibfnamefont {A.~W.}\ \bibnamefont {Ludwig}},\ }\bibfield  {title} {\bibinfo {title} {{M}ajorana fermions, exact mapping between quantum impurity fixed points with four bulk fermion species, and solution of the ``unitarity puzzle''},\ }\href {https://doi.org/https://doi.org/10.1016/S0550-3213(97)00596-8} {\bibfield  {journal} {\bibinfo  {journal} {Nucl. Phys. B}\ }\textbf {\bibinfo {volume} {506}},\ \bibinfo {pages} {565} (\bibinfo {year} {1997})}\BibitemShut {NoStop}%
\bibitem [{\citenamefont {Zhang}\ \emph {et~al.}(1999)\citenamefont {Zhang}, \citenamefont {Hewson},\ and\ \citenamefont {Bulla}}]{Zhang1999}%
  \BibitemOpen
  \bibfield  {author} {\bibinfo {author} {\bibfnamefont {G.-M.}\ \bibnamefont {Zhang}}, \bibinfo {author} {\bibfnamefont {A.}~\bibnamefont {Hewson}},\ and\ \bibinfo {author} {\bibfnamefont {R.}~\bibnamefont {Bulla}},\ }\bibfield  {title} {\bibinfo {title} {{M}ajorana fermion formulation of the two channel {K}ondo model},\ }\href {https://doi.org/https://doi.org/10.1016/S0038-1098(99)00294-X} {\bibfield  {journal} {\bibinfo  {journal} {Sol. State Comm.}\ }\textbf {\bibinfo {volume} {112}},\ \bibinfo {pages} {105} (\bibinfo {year} {1999})}\BibitemShut {NoStop}%
\bibitem [{Note1()}]{Note1}%
  \BibitemOpen
  \bibinfo {note} {A different construction starts from the ph-symmetric single-channel Anderson model, rewritten in terms of Majorana fermions, for then arbitrarily changing the hybridization term to break the $SO(4)$ symmetry down to $SO(3)$ \cite {Zhang1996,*Zhang1996a}. This gives a model that shares some features of the two-channel Kondo model, but does not start from a \protect \textit {bona fide} two-channel Anderson model \cite {bolech2006a,*iucci2008}. We shall leave aside those connections to be explored elsewhere.}\BibitemShut {Stop}%
\bibitem [{Note2()}]{Note2}%
  \BibitemOpen
  \bibinfo {note} {Connected to two separate level-1 \protect \textit {current algebras} that can, however, be mapped to each other via a staggered particle-hole transformation that mixes sectors with different boundary conditions \cite {Zhang1999}.}\BibitemShut {Stop}%
\bibitem [{\citenamefont {Affleck}(1990)}]{affleck1990}%
  \BibitemOpen
  \bibfield  {author} {\bibinfo {author} {\bibfnamefont {I.}~\bibnamefont {Affleck}},\ }\bibfield  {title} {\bibinfo {title} {A current-algebra approach to the {K}ondo effect},\ }\href {https://doi.org/10.1016/0550-3213(90)90440-O} {\bibfield  {journal} {\bibinfo  {journal} {Nucl. Phys. B}\ }\textbf {\bibinfo {volume} {336}},\ \bibinfo {pages} {517} (\bibinfo {year} {1990})}\BibitemShut {NoStop}%
\bibitem [{\citenamefont {Affleck}\ and\ \citenamefont {Ludwig}(1991)}]{affleck1991}%
  \BibitemOpen
  \bibfield  {author} {\bibinfo {author} {\bibfnamefont {I.}~\bibnamefont {Affleck}}\ and\ \bibinfo {author} {\bibfnamefont {A.~W.~W.}\ \bibnamefont {Ludwig}},\ }\bibfield  {title} {\bibinfo {title} {Critical-theory of overscreened {K}ondo fixed-points},\ }\href {https://doi.org/10.1016/0550-3213(91)90419-X} {\bibfield  {journal} {\bibinfo  {journal} {Nucl. Phys. B}\ }\textbf {\bibinfo {volume} {360}},\ \bibinfo {pages} {641} (\bibinfo {year} {1991})}\BibitemShut {NoStop}%
\bibitem [{\citenamefont {Affleck}\ \emph {et~al.}(1992)\citenamefont {Affleck}, \citenamefont {Ludwig}, \citenamefont {Pang},\ and\ \citenamefont {Cox}}]{Affleck1992}%
  \BibitemOpen
  \bibfield  {author} {\bibinfo {author} {\bibfnamefont {I.}~\bibnamefont {Affleck}}, \bibinfo {author} {\bibfnamefont {A.~W.~W.}\ \bibnamefont {Ludwig}}, \bibinfo {author} {\bibfnamefont {H.-B.}\ \bibnamefont {Pang}},\ and\ \bibinfo {author} {\bibfnamefont {D.~L.}\ \bibnamefont {Cox}},\ }\bibfield  {title} {\bibinfo {title} {Relevance of anisotropy in the multichannel {K}ondo effect: {C}omparison of conformal field theory and numerical renormalization-group results},\ }\href {https://doi.org/10.1103/PhysRevB.45.7918} {\bibfield  {journal} {\bibinfo  {journal} {Phys. Rev. B}\ }\textbf {\bibinfo {volume} {45}},\ \bibinfo {pages} {7918} (\bibinfo {year} {1992})}\BibitemShut {NoStop}%
\bibitem [{\citenamefont {Coleman}\ and\ \citenamefont {Schofield}(1995)}]{coleman1995}%
  \BibitemOpen
  \bibfield  {author} {\bibinfo {author} {\bibfnamefont {P.}~\bibnamefont {Coleman}}\ and\ \bibinfo {author} {\bibfnamefont {A.~J.}\ \bibnamefont {Schofield}},\ }\bibfield  {title} {\bibinfo {title} {Simple description of the anisotropic two-channel {K}ondo problem},\ }\href {https://doi.org/10.1103/PhysRevLett.75.2184} {\bibfield  {journal} {\bibinfo  {journal} {Phys. Rev. Lett.}\ }\textbf {\bibinfo {volume} {75}},\ \bibinfo {pages} {2184} (\bibinfo {year} {1995})}\BibitemShut {NoStop}%
\bibitem [{Note3()}]{Note3}%
  \BibitemOpen
  \bibinfo {note} {In the language of \protect \textit {boundary-condition-changing operators}, the impurity behaves differently in connection with the ``two-channel'' or ``compact'' bulks, respectively; cf.~Ref.~\protect \rev@citealp {affleck1994,*shah2003}.}\BibitemShut {Stop}%
\bibitem [{Note4()}]{Note4}%
  \BibitemOpen
  \bibinfo {note} {It was hoped that compactification would provide a blue print for stabilizing the non-Fermi-liquid physics against the otherwise-relevant channel asymmetry. This prospect found a match in the more recent proposals of topological Kondo impurities \cite {Beri2012,Altland2013,Altland2014,Li2023,*Koenig2023} (of high theoretical interest but facing steep experimental challenges) or of Kondo physics in correlated bulks \cite {Fiete2008,Koenig2020}, and in connection with more general $\protect \mathfrak {o}(N)$-Kondo models.}\BibitemShut {Stop}%
\bibitem [{\citenamefont {Bulla}\ and\ \citenamefont {Hewson}(1997)}]{bulla1997}%
  \BibitemOpen
  \bibfield  {author} {\bibinfo {author} {\bibfnamefont {R.}~\bibnamefont {Bulla}}\ and\ \bibinfo {author} {\bibfnamefont {A.~C.}\ \bibnamefont {Hewson}},\ }\bibfield  {title} {\bibinfo {title} {Numerical renormalization group study of the `compactified' {A}nderson model},\ }\href {https://doi.org/https://doi.org/10.1016/S0921-4526(96)00766-1} {\bibfield  {journal} {\bibinfo  {journal} {Physica B: Cond. Matt.}\ }\textbf {\bibinfo {volume} {230-232}},\ \bibinfo {pages} {627} (\bibinfo {year} {1997})}\BibitemShut {NoStop}%
\bibitem [{\citenamefont {Bulla}\ \emph {et~al.}(1997)\citenamefont {Bulla}, \citenamefont {Hewson},\ and\ \citenamefont {Zhang}}]{bulla1997a}%
  \BibitemOpen
  \bibfield  {author} {\bibinfo {author} {\bibfnamefont {R.}~\bibnamefont {Bulla}}, \bibinfo {author} {\bibfnamefont {A.~C.}\ \bibnamefont {Hewson}},\ and\ \bibinfo {author} {\bibfnamefont {G.-M.}\ \bibnamefont {Zhang}},\ }\bibfield  {title} {\bibinfo {title} {Low-energy fixed points of the \ensuremath{\sigma}-\ensuremath{\tau} and the {$O(3)$} symmetric {A}nderson models},\ }\href {https://doi.org/10.1103/PhysRevB.56.11721} {\bibfield  {journal} {\bibinfo  {journal} {Phys. Rev. B}\ }\textbf {\bibinfo {volume} {56}},\ \bibinfo {pages} {11721} (\bibinfo {year} {1997})}\BibitemShut {NoStop}%
\bibitem [{\citenamefont {Ye}(1998)}]{ye1998}%
  \BibitemOpen
  \bibfield  {author} {\bibinfo {author} {\bibfnamefont {J.}~\bibnamefont {Ye}},\ }\bibfield  {title} {\bibinfo {title} {On two channel flavor anisotropic and one channel compactified {K}ondo models},\ }\href {https://doi.org/https://doi.org/10.1016/S0550-3213(97)80485-3} {\bibfield  {journal} {\bibinfo  {journal} {Nucl. Phys. B}\ }\textbf {\bibinfo {volume} {512}},\ \bibinfo {pages} {543} (\bibinfo {year} {1998})}\BibitemShut {NoStop}%
\bibitem [{\citenamefont {Schofield}(1997)}]{schofield1997}%
  \BibitemOpen
  \bibfield  {author} {\bibinfo {author} {\bibfnamefont {A.~J.}\ \bibnamefont {Schofield}},\ }\bibfield  {title} {\bibinfo {title} {Bosonization in the two-channel {K}ondo model},\ }\href {https://doi.org/10.1103/PhysRevB.55.5627} {\bibfield  {journal} {\bibinfo  {journal} {Phys. Rev. B}\ }\textbf {\bibinfo {volume} {55}},\ \bibinfo {pages} {5627} (\bibinfo {year} {1997})},\ \bibinfo {note} {cf.~arXiv:cond-mat/9606063 for mismatched equation numbers.}\BibitemShut {Stop}%
\bibitem [{\citenamefont {Ljepoja}\ \emph {et~al.}(2024{\natexlab{a}})\citenamefont {Ljepoja}, \citenamefont {Bolech},\ and\ \citenamefont {Shah}}]{Ljepoja2024a}%
  \BibitemOpen
  \bibfield  {author} {\bibinfo {author} {\bibfnamefont {A.}~\bibnamefont {Ljepoja}}, \bibinfo {author} {\bibfnamefont {C.~J.}\ \bibnamefont {Bolech}},\ and\ \bibinfo {author} {\bibfnamefont {N.}~\bibnamefont {Shah}},\ }\bibfield  {title} {\bibinfo {title} {Systematic compactification of the two-channel {K}ondo model. {I}. {C}onsistent bosonization-debosonization approach and exact comparisons},\ }\href {https://doi.org/10.1103/PhysRevB.110.045108} {\bibfield  {journal} {\bibinfo  {journal} {Phys. Rev. B}\ }\textbf {\bibinfo {volume} {110}},\ \bibinfo {pages} {045108} (\bibinfo {year} {2024}{\natexlab{a}})}\BibitemShut {NoStop}%
\bibitem [{\citenamefont {Shah}\ and\ \citenamefont {Bolech}(2016)}]{shah2016}%
  \BibitemOpen
  \bibfield  {author} {\bibinfo {author} {\bibfnamefont {N.}~\bibnamefont {Shah}}\ and\ \bibinfo {author} {\bibfnamefont {C.~J.}\ \bibnamefont {Bolech}},\ }\bibfield  {title} {\bibinfo {title} {Consistent bosonization-debosonization. {I}. {A} resolution of the nonequilibrium transport puzzle},\ }\href {https://doi.org/10.1103/PhysRevB.93.085440} {\bibfield  {journal} {\bibinfo  {journal} {Phys. Rev. B}\ }\textbf {\bibinfo {volume} {93}},\ \bibinfo {pages} {085440} (\bibinfo {year} {2016})}\BibitemShut {NoStop}%
\bibitem [{\citenamefont {Bolech}\ and\ \citenamefont {Shah}(2016)}]{bolech2016}%
  \BibitemOpen
  \bibfield  {author} {\bibinfo {author} {\bibfnamefont {C.~J.}\ \bibnamefont {Bolech}}\ and\ \bibinfo {author} {\bibfnamefont {N.}~\bibnamefont {Shah}},\ }\bibfield  {title} {\bibinfo {title} {Consistent bosonization-debosonization. {II}. {T}he two-lead {K}ondo problem and the fate of its nonequilibrium {T}oulouse point},\ }\href {https://doi.org/10.1103/PhysRevB.93.085441} {\bibfield  {journal} {\bibinfo  {journal} {Phys. Rev. B}\ }\textbf {\bibinfo {volume} {93}},\ \bibinfo {pages} {085441} (\bibinfo {year} {2016})}\BibitemShut {NoStop}%
\bibitem [{\citenamefont {Wilson}(1975)}]{wilson1975}%
  \BibitemOpen
  \bibfield  {author} {\bibinfo {author} {\bibfnamefont {K.~G.}\ \bibnamefont {Wilson}},\ }\bibfield  {title} {\bibinfo {title} {The renormalization group: {C}ritical phenomena and the {K}ondo problem},\ }\href {https://doi.org/10.1103/RevModPhys.47.773} {\bibfield  {journal} {\bibinfo  {journal} {Rev. Mod. Phys.}\ }\textbf {\bibinfo {volume} {47}},\ \bibinfo {pages} {773} (\bibinfo {year} {1975})}\BibitemShut {NoStop}%
\bibitem [{\citenamefont {Andrei}\ \emph {et~al.}(1983)\citenamefont {Andrei}, \citenamefont {Furuya},\ and\ \citenamefont {Lowenstein}}]{andrei1983}%
  \BibitemOpen
  \bibfield  {author} {\bibinfo {author} {\bibfnamefont {N.}~\bibnamefont {Andrei}}, \bibinfo {author} {\bibfnamefont {K.}~\bibnamefont {Furuya}},\ and\ \bibinfo {author} {\bibfnamefont {J.~H.}\ \bibnamefont {Lowenstein}},\ }\bibfield  {title} {\bibinfo {title} {Solution of the {K}ondo problem},\ }\href {https://doi.org/10.1103/RevModPhys.55.331} {\bibfield  {journal} {\bibinfo  {journal} {Rev. Mod. Phys.}\ }\textbf {\bibinfo {volume} {55}},\ \bibinfo {pages} {331} (\bibinfo {year} {1983})}\BibitemShut {NoStop}%
\bibitem [{\citenamefont {Anderson}(1970)}]{Anderson1970}%
  \BibitemOpen
  \bibfield  {author} {\bibinfo {author} {\bibfnamefont {P.~W.}\ \bibnamefont {Anderson}},\ }\bibfield  {title} {\bibinfo {title} {A poor man's derivation of scaling laws for the {K}ondo problem},\ }\href {https://doi.org/10.1088/0022-3719/3/12/008} {\bibfield  {journal} {\bibinfo  {journal} {J. Phys. C: Solid St. Phys.}\ }\textbf {\bibinfo {volume} {3}},\ \bibinfo {pages} {2436} (\bibinfo {year} {1970})}\BibitemShut {NoStop}%
\bibitem [{\citenamefont {Wegner}(1972)}]{Wegner1972}%
  \BibitemOpen
  \bibfield  {author} {\bibinfo {author} {\bibfnamefont {F.~J.}\ \bibnamefont {Wegner}},\ }\bibfield  {title} {\bibinfo {title} {Corrections to scaling laws},\ }\href {https://doi.org/10.1103/PhysRevB.5.4529} {\bibfield  {journal} {\bibinfo  {journal} {Phys. Rev. B}\ }\textbf {\bibinfo {volume} {5}},\ \bibinfo {pages} {4529} (\bibinfo {year} {1972})}\BibitemShut {NoStop}%
\bibitem [{\citenamefont {Wegner}\ and\ \citenamefont {Houghton}(1973)}]{Wegner1973}%
  \BibitemOpen
  \bibfield  {author} {\bibinfo {author} {\bibfnamefont {F.~J.}\ \bibnamefont {Wegner}}\ and\ \bibinfo {author} {\bibfnamefont {A.}~\bibnamefont {Houghton}},\ }\bibfield  {title} {\bibinfo {title} {Renormalization group equation for critical phenomena},\ }\href {https://doi.org/10.1103/PhysRevA.8.401} {\bibfield  {journal} {\bibinfo  {journal} {Phys. Rev. A}\ }\textbf {\bibinfo {volume} {8}},\ \bibinfo {pages} {401} (\bibinfo {year} {1973})}\BibitemShut {NoStop}%
\bibitem [{\citenamefont {Wilson}\ and\ \citenamefont {Kogut}(1974)}]{Wilson1974}%
  \BibitemOpen
  \bibfield  {author} {\bibinfo {author} {\bibfnamefont {K.~G.}\ \bibnamefont {Wilson}}\ and\ \bibinfo {author} {\bibfnamefont {J.}~\bibnamefont {Kogut}},\ }\bibfield  {title} {\bibinfo {title} {The renormalization group and the $\epsilon$ expansion},\ }\href {https://doi.org/https://doi.org/10.1016/0370-1573(74)90023-4} {\bibfield  {journal} {\bibinfo  {journal} {Phys. Rep.}\ }\textbf {\bibinfo {volume} {12}},\ \bibinfo {pages} {75} (\bibinfo {year} {1974})}\BibitemShut {NoStop}%
\bibitem [{\citenamefont {Hewson}(1993)}]{Hewson}%
  \BibitemOpen
  \bibfield  {author} {\bibinfo {author} {\bibfnamefont {A.~C.}\ \bibnamefont {Hewson}},\ }\href {https://doi.org/10.1017/CBO9780511470752} {\emph {\bibinfo {title} {The {K}ondo Problem to {H}eavy {F}ermions}}}\ (\bibinfo  {publisher} {Cambridge University Press},\ \bibinfo {address} {Cambridge UK},\ \bibinfo {year} {1993})\BibitemShut {NoStop}%
\bibitem [{\citenamefont {Nevidomskyy}(2015)}]{Nevidomskyy2015}%
  \BibitemOpen
  \bibfield  {author} {\bibinfo {author} {\bibfnamefont {A.~H.}\ \bibnamefont {Nevidomskyy}},\ }\bibfield  {title} {\bibinfo {title} {The {K}ondo model and poor man’s scaling},\ }in\ \href@noop {} {\emph {\bibinfo {booktitle} {Many-Body Physics: From {K}ondo to {H}ubbard}}},\ \bibinfo {series} {Modeling and Simulation}, Vol.~\bibinfo {volume} {5},\ \bibinfo {editor} {edited by\ \bibinfo {editor} {\bibfnamefont {E.}~\bibnamefont {Pavarini}}, \bibinfo {editor} {\bibfnamefont {E.}~\bibnamefont {Koch}},\ and\ \bibinfo {editor} {\bibfnamefont {P.}~\bibnamefont {Coleman}}}\ (\bibinfo {address} {Forschungszentrum Julich, Germany},\ \bibinfo {year} {2015})\ Chap.~\bibinfo {chapter} {4}\BibitemShut {NoStop}%
\bibitem [{\citenamefont {Kuramoto}(1998)}]{kuramoto1998}%
  \BibitemOpen
  \bibfield  {author} {\bibinfo {author} {\bibfnamefont {Y.}~\bibnamefont {Kuramoto}},\ }\bibfield  {title} {\bibinfo {title} {Perturbative renormalization of multi-channel {K}ondo-type models},\ }\href {https://doi.org/10.1007/s100510050466} {\bibfield  {journal} {\bibinfo  {journal} {Euro. Phys. J. B}\ }\textbf {\bibinfo {volume} {5}},\ \bibinfo {pages} {457} (\bibinfo {year} {1998})}\BibitemShut {NoStop}%
\bibitem [{\citenamefont {S\'olyom}\ and\ \citenamefont {Zawadoswki}(1974)}]{Solyom1974}%
  \BibitemOpen
  \bibfield  {author} {\bibinfo {author} {\bibfnamefont {J.}~\bibnamefont {S\'olyom}}\ and\ \bibinfo {author} {\bibfnamefont {A.}~\bibnamefont {Zawadoswki}},\ }\bibfield  {title} {\bibinfo {title} {Are the scaling laws for the {K}ondo problem exact?},\ }\href {https://doi.org/10.1088/0305-4608/4/1/009} {\bibfield  {journal} {\bibinfo  {journal} {J. Phys. F: Met. Phys.}\ }\textbf {\bibinfo {volume} {4}},\ \bibinfo {pages} {80} (\bibinfo {year} {1974})}\BibitemShut {NoStop}%
\bibitem [{\citenamefont {Gan}\ \emph {et~al.}(1993)\citenamefont {Gan}, \citenamefont {Andrei},\ and\ \citenamefont {Coleman}}]{gan1993}%
  \BibitemOpen
  \bibfield  {author} {\bibinfo {author} {\bibfnamefont {J.}~\bibnamefont {Gan}}, \bibinfo {author} {\bibfnamefont {N.}~\bibnamefont {Andrei}},\ and\ \bibinfo {author} {\bibfnamefont {P.}~\bibnamefont {Coleman}},\ }\bibfield  {title} {\bibinfo {title} {Perturbative approach to the non-{F}ermi-liquid fixed point of the overscreened {K}ondo problem},\ }\href {https://doi.org/10.1103/PhysRevLett.70.686} {\bibfield  {journal} {\bibinfo  {journal} {Phys. Rev. Lett.}\ }\textbf {\bibinfo {volume} {70}},\ \bibinfo {pages} {686} (\bibinfo {year} {1993})}\BibitemShut {NoStop}%
\bibitem [{\citenamefont {Gan}(1994)}]{gan1994}%
  \BibitemOpen
  \bibfield  {author} {\bibinfo {author} {\bibfnamefont {J.}~\bibnamefont {Gan}},\ }\bibfield  {title} {\bibinfo {title} {On the multichannel {K}ondo model},\ }\href {https://doi.org/10.1088/0953-8984/6/24/016} {\bibfield  {journal} {\bibinfo  {journal} {J. Phys.: Cond. Mat.}\ }\textbf {\bibinfo {volume} {6}},\ \bibinfo {pages} {4547} (\bibinfo {year} {1994})}\BibitemShut {NoStop}%
\bibitem [{\citenamefont {Fischer}(1999)}]{fischer1999}%
  \BibitemOpen
  \bibfield  {author} {\bibinfo {author} {\bibfnamefont {K.}~\bibnamefont {Fischer}},\ }\bibfield  {title} {\bibinfo {title} {$\ensuremath{\beta}$ function of the multichannel {K}ondo model},\ }\href {https://doi.org/10.1103/PhysRevB.59.14845} {\bibfield  {journal} {\bibinfo  {journal} {Phys. Rev. B}\ }\textbf {\bibinfo {volume} {59}},\ \bibinfo {pages} {14845} (\bibinfo {year} {1999})}\BibitemShut {NoStop}%
\bibitem [{\citenamefont {Ljepoja}\ \emph {et~al.}(2024{\natexlab{b}})\citenamefont {Ljepoja}, \citenamefont {Bolech},\ and\ \citenamefont {Shah}}]{Ljepoja2024c}%
  \BibitemOpen
  \bibfield  {author} {\bibinfo {author} {\bibfnamefont {A.}~\bibnamefont {Ljepoja}}, \bibinfo {author} {\bibfnamefont {C.~J.}\ \bibnamefont {Bolech}},\ and\ \bibinfo {author} {\bibfnamefont {N.}~\bibnamefont {Shah}},\ }\bibfield  {title} {\bibinfo {title} {Systematic compactification of the two-channel {K}ondo model. {III}. {E}xtended field-theoretic renormalization group analysis},\ }\href {https://doi.org/10.1103/PhysRevB.110.045110} {\bibfield  {journal} {\bibinfo  {journal} {Phys. Rev. B}\ }\textbf {\bibinfo {volume} {110}},\ \bibinfo {pages} {045110} (\bibinfo {year} {2024}{\natexlab{b}})}\BibitemShut {NoStop}%
\bibitem [{\citenamefont {Affleck}(2005)}]{affleck2005}%
  \BibitemOpen
  \bibfield  {author} {\bibinfo {author} {\bibfnamefont {I.}~\bibnamefont {Affleck}},\ }\bibfield  {title} {\bibinfo {title} {Non-{F}ermi liquid behavior in {K}ondo models},\ }\href {https://doi.org/10.1143/JPSJ.74.59} {\bibfield  {journal} {\bibinfo  {journal} {J. Phys. Soc. Jpn.}\ }\textbf {\bibinfo {volume} {74}},\ \bibinfo {pages} {59} (\bibinfo {year} {2005})}\BibitemShut {NoStop}%
\bibitem [{Note5()}]{Note5}%
  \BibitemOpen
  \bibinfo {note} {It is important to notice that diagrams of the type of that in Fig.~\ref {fig:4th} are the only ones that contribute to the channel-independent part of the scaling, since they are the only irreducible fourth-order diagrams without fermionic loops. However, besides the $J_{\perp }^4$ unphysical contribution to the flow of $J_{z}$, there is also a $J_{\perp }^3 J_{z}$ contribution to the flow of $J_{\perp }$ which is also unphysical and untranslatable; and those are the only two untranslatable contributions at this order of perturbation (all the others are physical instances of this diagram that have a correlate in the direct calculation of the beta function; cf.~Appendix B of Ref.~\protect \rev@citealp {affleck1991}).}\BibitemShut {Stop}%
\bibitem [{Note6()}]{Note6}%
  \BibitemOpen
  \bibinfo {note} {The beta-function coefficient $a$ will have two parts: a channel-dependent one, which will be the same in all schemes and given by $(1+\protect \mathrm {\ln (2)})M$, plus a channel-independent part that will differ between the schemes, (the direct-scheme calculation will give $\pi ^2 / 4$, whereas the alternative value will be $11 \pi ^2 / 64$ for the conventional-scheme).}\BibitemShut {Stop}%
\bibitem [{\citenamefont {Barzykin}\ and\ \citenamefont {Affleck}(1998)}]{Barzykin1998}%
  \BibitemOpen
  \bibfield  {author} {\bibinfo {author} {\bibfnamefont {V.}~\bibnamefont {Barzykin}}\ and\ \bibinfo {author} {\bibfnamefont {I.}~\bibnamefont {Affleck}},\ }\bibfield  {title} {\bibinfo {title} {Screening cloud in the $k$-channel {K}ondo model: {P}erturbative and large-$k$ results},\ }\href {https://doi.org/10.1103/PhysRevB.57.432} {\bibfield  {journal} {\bibinfo  {journal} {Phys. Rev. B}\ }\textbf {\bibinfo {volume} {57}},\ \bibinfo {pages} {432} (\bibinfo {year} {1998})},\ \bibinfo {note} {erratum: Phys. Rev. B \textbf{75}, 019906 (2007)}\BibitemShut {NoStop}%
\bibitem [{Note7()}]{Note7}%
  \BibitemOpen
  \bibinfo {note} {One could also more simply adopt $g^\star /2$, the mid point between the two fixed points. Such an alternative convention would result in the same leading exponential behavior, $\exp (-g^\star /g_0\Delta )$, and slightly simpler numerical factors, but is less well physically motivated.}\BibitemShut {Stop}%
\bibitem [{\citenamefont {Appelquist}\ \emph {et~al.}(1998)\citenamefont {Appelquist}, \citenamefont {Ratnaweera}, \citenamefont {Terning},\ and\ \citenamefont {Wijewardhana}}]{Appelquist1998}%
  \BibitemOpen
  \bibfield  {author} {\bibinfo {author} {\bibfnamefont {T.}~\bibnamefont {Appelquist}}, \bibinfo {author} {\bibfnamefont {A.}~\bibnamefont {Ratnaweera}}, \bibinfo {author} {\bibfnamefont {J.}~\bibnamefont {Terning}},\ and\ \bibinfo {author} {\bibfnamefont {L.~C.~R.}\ \bibnamefont {Wijewardhana}},\ }\bibfield  {title} {\bibinfo {title} {Phase structure of an {$\mathrm{SU}(N)$} gauge theory with ${N}_{f}$ flavors},\ }\href {https://doi.org/10.1103/PhysRevD.58.105017} {\bibfield  {journal} {\bibinfo  {journal} {Phys. Rev. D}\ }\textbf {\bibinfo {volume} {58}},\ \bibinfo {pages} {105017} (\bibinfo {year} {1998})}\BibitemShut {NoStop}%
\bibitem [{\citenamefont {Kondo}(1968)}]{Kondo1968}%
  \BibitemOpen
  \bibfield  {author} {\bibinfo {author} {\bibfnamefont {J.}~\bibnamefont {Kondo}},\ }\bibfield  {title} {\bibinfo {title} {Free-energy shift of conduction electrons due to the $s$-$d$ exchange interaction},\ }\href {https://doi.org/10.1143/PTP.40.683} {\bibfield  {journal} {\bibinfo  {journal} {Prog. Theor. Phys.}\ }\textbf {\bibinfo {volume} {40}},\ \bibinfo {pages} {683} (\bibinfo {year} {1968})}\BibitemShut {NoStop}%
\bibitem [{\citenamefont {Bensimon}\ \emph {et~al.}(2006)\citenamefont {Bensimon}, \citenamefont {Jerez},\ and\ \citenamefont {Lavagna}}]{bensimon2006}%
  \BibitemOpen
  \bibfield  {author} {\bibinfo {author} {\bibfnamefont {D.}~\bibnamefont {Bensimon}}, \bibinfo {author} {\bibfnamefont {A.}~\bibnamefont {Jerez}},\ and\ \bibinfo {author} {\bibfnamefont {M.}~\bibnamefont {Lavagna}},\ }\bibfield  {title} {\bibinfo {title} {Intermediate coupling fixed point study in the overscreened regime of generalized multichannel $\mathrm{SU}({N})$ {K}ondo models},\ }\href {https://doi.org/10.1103/PhysRevB.73.224445} {\bibfield  {journal} {\bibinfo  {journal} {Phys. Rev. B}\ }\textbf {\bibinfo {volume} {73}},\ \bibinfo {pages} {224445} (\bibinfo {year} {2006})}\BibitemShut {NoStop}%
\bibitem [{\citenamefont {Andrei}\ and\ \citenamefont {Destri}(1984)}]{andrei1984}%
  \BibitemOpen
  \bibfield  {author} {\bibinfo {author} {\bibfnamefont {N.}~\bibnamefont {Andrei}}\ and\ \bibinfo {author} {\bibfnamefont {C.}~\bibnamefont {Destri}},\ }\bibfield  {title} {\bibinfo {title} {Solution of the multichannel {K}ondo problem},\ }\href {https://doi.org/10.1103/PhysRevLett.52.364} {\bibfield  {journal} {\bibinfo  {journal} {Phys. Rev. Lett.}\ }\textbf {\bibinfo {volume} {52}},\ \bibinfo {pages} {364} (\bibinfo {year} {1984})}\BibitemShut {NoStop}%
\bibitem [{\citenamefont {Bolech}\ and\ \citenamefont {Andrei}(2002)}]{bolech2002}%
  \BibitemOpen
  \bibfield  {author} {\bibinfo {author} {\bibfnamefont {C.~J.}\ \bibnamefont {Bolech}}\ and\ \bibinfo {author} {\bibfnamefont {N.}~\bibnamefont {Andrei}},\ }\bibfield  {title} {\bibinfo {title} {Solution of the two-channel {A}nderson impurity model: Implications for the heavy fermion {UBe$_{13}$}},\ }\href {https://doi.org/10.1103/PhysRevLett.88.237206} {\bibfield  {journal} {\bibinfo  {journal} {Phys. Rev. Lett.}\ }\textbf {\bibinfo {volume} {88}},\ \bibinfo {pages} {237206} (\bibinfo {year} {2002})}\BibitemShut {NoStop}%
\bibitem [{\citenamefont {Bolech}\ and\ \citenamefont {Andrei}(2005)}]{bolech2005a}%
  \BibitemOpen
  \bibfield  {author} {\bibinfo {author} {\bibfnamefont {C.~J.}\ \bibnamefont {Bolech}}\ and\ \bibinfo {author} {\bibfnamefont {N.}~\bibnamefont {Andrei}},\ }\bibfield  {title} {\bibinfo {title} {Solution of the multi-channel {A}nderson impurity model: {G}round state and thermodynamics},\ }\href {https://doi.org/10.1103/PhysRevB.71.205104} {\bibfield  {journal} {\bibinfo  {journal} {Phys. Rev. B}\ }\textbf {\bibinfo {volume} {71}},\ \bibinfo {pages} {205104} (\bibinfo {year} {2005})}\BibitemShut {NoStop}%
\bibitem [{\citenamefont {Johannesson}\ \emph {et~al.}(2003)\citenamefont {Johannesson}, \citenamefont {Andrei},\ and\ \citenamefont {Bolech}}]{johannesson2003}%
  \BibitemOpen
  \bibfield  {author} {\bibinfo {author} {\bibfnamefont {H.}~\bibnamefont {Johannesson}}, \bibinfo {author} {\bibfnamefont {N.}~\bibnamefont {Andrei}},\ and\ \bibinfo {author} {\bibfnamefont {C.~J.}\ \bibnamefont {Bolech}},\ }\bibfield  {title} {\bibinfo {title} {Critical theory of the two-channel {A}nderson impurity model},\ }\href {https://doi.org/10.1103/PhysRevB.68.075112} {\bibfield  {journal} {\bibinfo  {journal} {Phys. Rev. B}\ }\textbf {\bibinfo {volume} {68}},\ \bibinfo {pages} {075112} (\bibinfo {year} {2003})}\BibitemShut {NoStop}%
\bibitem [{\citenamefont {Johannesson}\ \emph {et~al.}(2005)\citenamefont {Johannesson}, \citenamefont {Bolech},\ and\ \citenamefont {Andrei}}]{johannesson2005}%
  \BibitemOpen
  \bibfield  {author} {\bibinfo {author} {\bibfnamefont {H.}~\bibnamefont {Johannesson}}, \bibinfo {author} {\bibfnamefont {C.~J.}\ \bibnamefont {Bolech}},\ and\ \bibinfo {author} {\bibfnamefont {N.}~\bibnamefont {Andrei}},\ }\bibfield  {title} {\bibinfo {title} {Two-channel {A}nderson impurity model: {S}ingle-electron {G}reen's function, self-energies, and resistivity},\ }\href {https://doi.org/10.1103/PhysRevB.71.195107} {\bibfield  {journal} {\bibinfo  {journal} {Phys. Rev. B}\ }\textbf {\bibinfo {volume} {71}},\ \bibinfo {pages} {195107} (\bibinfo {year} {2005})}\BibitemShut {NoStop}%
\bibitem [{\citenamefont {Oreg}\ and\ \citenamefont {Goldhaber-Gordon}(2003)}]{oreg2003}%
  \BibitemOpen
  \bibfield  {author} {\bibinfo {author} {\bibfnamefont {Y.}~\bibnamefont {Oreg}}\ and\ \bibinfo {author} {\bibfnamefont {D.}~\bibnamefont {Goldhaber-Gordon}},\ }\bibfield  {title} {\bibinfo {title} {Two-channel {K}ondo effect in a modified single electron transistor},\ }\href {https://doi.org/10.1103/PhysRevLett.90.136602} {\bibfield  {journal} {\bibinfo  {journal} {Phys. Rev. Lett.}\ }\textbf {\bibinfo {volume} {90}},\ \bibinfo {pages} {136602} (\bibinfo {year} {2003})}\BibitemShut {NoStop}%
\bibitem [{\citenamefont {Potok}\ \emph {et~al.}(2007)\citenamefont {Potok}, \citenamefont {Rau}, \citenamefont {Shtrikman}, \citenamefont {Oreg},\ and\ \citenamefont {Goldhaber-Gordon}}]{potok2007}%
  \BibitemOpen
  \bibfield  {author} {\bibinfo {author} {\bibfnamefont {R.~M.}\ \bibnamefont {Potok}}, \bibinfo {author} {\bibfnamefont {I.~G.}\ \bibnamefont {Rau}}, \bibinfo {author} {\bibfnamefont {H.}~\bibnamefont {Shtrikman}}, \bibinfo {author} {\bibfnamefont {Y.}~\bibnamefont {Oreg}},\ and\ \bibinfo {author} {\bibfnamefont {D.}~\bibnamefont {Goldhaber-Gordon}},\ }\bibfield  {title} {\bibinfo {title} {Observation of the two-channel {K}ondo effect},\ }\href {https://doi.org/10.1038/nature05556} {\bibfield  {journal} {\bibinfo  {journal} {Nature}\ }\textbf {\bibinfo {volume} {446}},\ \bibinfo {pages} {167} (\bibinfo {year} {2007})}\BibitemShut {NoStop}%
\bibitem [{\citenamefont {Bolech}\ and\ \citenamefont {Shah}(2005)}]{bolech2005b}%
  \BibitemOpen
  \bibfield  {author} {\bibinfo {author} {\bibfnamefont {C.~J.}\ \bibnamefont {Bolech}}\ and\ \bibinfo {author} {\bibfnamefont {N.}~\bibnamefont {Shah}},\ }\bibfield  {title} {\bibinfo {title} {Prediction of the capacitance line shape in two-channel quantum dots},\ }\href {https://doi.org/10.1103/PhysRevLett.95.036801} {\bibfield  {journal} {\bibinfo  {journal} {Phys. Rev. Lett.}\ }\textbf {\bibinfo {volume} {95}},\ \bibinfo {pages} {036801} (\bibinfo {year} {2005})}\BibitemShut {NoStop}%
\bibitem [{\citenamefont {Shah}\ and\ \citenamefont {Rosch}(2006)}]{shah2006}%
  \BibitemOpen
  \bibfield  {author} {\bibinfo {author} {\bibfnamefont {N.}~\bibnamefont {Shah}}\ and\ \bibinfo {author} {\bibfnamefont {A.}~\bibnamefont {Rosch}},\ }\bibfield  {title} {\bibinfo {title} {{N}onequilibrium conductance of a three-terminal quantum dot in the {K}ondo regime: {P}erturbative renormalization group study},\ }\href {https://doi.org/10.1103/PhysRevB.73.081309} {\bibfield  {journal} {\bibinfo  {journal} {Phys. Rev. B}\ }\textbf {\bibinfo {volume} {73}},\ \bibinfo {pages} {081309} (\bibinfo {year} {2006})}\BibitemShut {NoStop}%
\bibitem [{\citenamefont {Mitra}\ and\ \citenamefont {Rosch}(2011)}]{mitra2011}%
  \BibitemOpen
  \bibfield  {author} {\bibinfo {author} {\bibfnamefont {A.}~\bibnamefont {Mitra}}\ and\ \bibinfo {author} {\bibfnamefont {A.}~\bibnamefont {Rosch}},\ }\bibfield  {title} {\bibinfo {title} {Current-induced decoherence in the multichannel {K}ondo problem},\ }\href {https://doi.org/10.1103/PhysRevLett.106.106402} {\bibfield  {journal} {\bibinfo  {journal} {Phys. Rev. Lett.}\ }\textbf {\bibinfo {volume} {106}},\ \bibinfo {pages} {106402} (\bibinfo {year} {2011})}\BibitemShut {NoStop}%
\bibitem [{Note8()}]{Note8}%
  \BibitemOpen
  \bibinfo {note} {For simplicity, we will not add a co-tunneling term. In that way the model reduces, at the channel-symmetric point, to the one we have been studying so far, and, moreover, the setting is similar to that of the so-called ``charge Kondo'' model \cite {matveev1991,*matveev1995,*Furusaki1995,Iftikhar2015,*Iftikhar2018,Karki2022,Pouse2023}, which was used to experimentally demonstrate the RG flow with channel asymmetry.}\BibitemShut {Stop}%
\bibitem [{\citenamefont {Bolech}\ and\ \citenamefont {Iucci}(2006)}]{bolech2006a}%
  \BibitemOpen
  \bibfield  {author} {\bibinfo {author} {\bibfnamefont {C.~J.}\ \bibnamefont {Bolech}}\ and\ \bibinfo {author} {\bibfnamefont {A.}~\bibnamefont {Iucci}},\ }\bibfield  {title} {\bibinfo {title} {Mapping of the anisotropic two-channel {A}nderson model onto a {F}ermi-{M}ajorana biresonant level model},\ }\href {https://doi.org/10.1103/PhysRevLett.95.056402} {\bibfield  {journal} {\bibinfo  {journal} {Phys. Rev. Lett.}\ }\textbf {\bibinfo {volume} {96}},\ \bibinfo {pages} {056402} (\bibinfo {year} {2006})}\BibitemShut {NoStop}%
\bibitem [{\citenamefont {Iucci}\ and\ \citenamefont {Bolech}(2008)}]{iucci2008}%
  \BibitemOpen
  \bibfield  {author} {\bibinfo {author} {\bibfnamefont {A.}~\bibnamefont {Iucci}}\ and\ \bibinfo {author} {\bibfnamefont {C.~J.}\ \bibnamefont {Bolech}},\ }\bibfield  {title} {\bibinfo {title} {Bosonization approach to the mixed-valence two-channel {K}ondo problem},\ }\href {https://doi.org/10.1103/PhysRevB.77.195113} {\bibfield  {journal} {\bibinfo  {journal} {Phys. Rev. B}\ }\textbf {\bibinfo {volume} {77}},\ \bibinfo {pages} {195113} (\bibinfo {year} {2008})}\BibitemShut {NoStop}%
\bibitem [{\citenamefont {Affleck}\ and\ \citenamefont {Ludwig}(1994)}]{affleck1994}%
  \BibitemOpen
  \bibfield  {author} {\bibinfo {author} {\bibfnamefont {I.}~\bibnamefont {Affleck}}\ and\ \bibinfo {author} {\bibfnamefont {A.~W.~W.}\ \bibnamefont {Ludwig}},\ }\bibfield  {title} {\bibinfo {title} {The {F}ermi edge singularity and boundary condition changing operators},\ }\href {https://doi.org/10.1088/0305-4470/27/16/007} {\bibfield  {journal} {\bibinfo  {journal} {J. Phys. A: Math. Gen.}\ }\textbf {\bibinfo {volume} {27}},\ \bibinfo {pages} {5375} (\bibinfo {year} {1994})}\BibitemShut {NoStop}%
\bibitem [{\citenamefont {Shah}\ and\ \citenamefont {Millis}(2003)}]{shah2003}%
  \BibitemOpen
  \bibfield  {author} {\bibinfo {author} {\bibfnamefont {N.}~\bibnamefont {Shah}}\ and\ \bibinfo {author} {\bibfnamefont {A.~J.}\ \bibnamefont {Millis}},\ }\bibfield  {title} {\bibinfo {title} {Dissipative dynamics of an extended magnetic nanostructure: {S}pin necklace in a metallic environment},\ }\href {https://doi.org/10.1103/PhysRevLett.91.147204} {\bibfield  {journal} {\bibinfo  {journal} {Phys. Rev. Lett.}\ }\textbf {\bibinfo {volume} {91}},\ \bibinfo {pages} {147204} (\bibinfo {year} {2003})}\BibitemShut {NoStop}%
\bibitem [{\citenamefont {B\'eri}\ and\ \citenamefont {Cooper}(2012)}]{Beri2012}%
  \BibitemOpen
  \bibfield  {author} {\bibinfo {author} {\bibfnamefont {B.}~\bibnamefont {B\'eri}}\ and\ \bibinfo {author} {\bibfnamefont {N.~R.}\ \bibnamefont {Cooper}},\ }\bibfield  {title} {\bibinfo {title} {Topological {K}ondo effect with {M}ajorana fermions},\ }\href {https://doi.org/10.1103/PhysRevLett.109.156803} {\bibfield  {journal} {\bibinfo  {journal} {Phys. Rev. Lett.}\ }\textbf {\bibinfo {volume} {109}},\ \bibinfo {pages} {156803} (\bibinfo {year} {2012})}\BibitemShut {NoStop}%
\bibitem [{\citenamefont {Altland}\ and\ \citenamefont {Egger}(2013)}]{Altland2013}%
  \BibitemOpen
  \bibfield  {author} {\bibinfo {author} {\bibfnamefont {A.}~\bibnamefont {Altland}}\ and\ \bibinfo {author} {\bibfnamefont {R.}~\bibnamefont {Egger}},\ }\bibfield  {title} {\bibinfo {title} {Multiterminal {C}oulomb-{M}ajorana junction},\ }\href {https://doi.org/10.1103/PhysRevLett.110.196401} {\bibfield  {journal} {\bibinfo  {journal} {Phys. Rev. Lett.}\ }\textbf {\bibinfo {volume} {110}},\ \bibinfo {pages} {196401} (\bibinfo {year} {2013})}\BibitemShut {NoStop}%
\bibitem [{\citenamefont {Altland}\ \emph {et~al.}(2014)\citenamefont {Altland}, \citenamefont {B\'eri}, \citenamefont {Egger},\ and\ \citenamefont {Tsvelik}}]{Altland2014}%
  \BibitemOpen
  \bibfield  {author} {\bibinfo {author} {\bibfnamefont {A.}~\bibnamefont {Altland}}, \bibinfo {author} {\bibfnamefont {B.}~\bibnamefont {B\'eri}}, \bibinfo {author} {\bibfnamefont {R.}~\bibnamefont {Egger}},\ and\ \bibinfo {author} {\bibfnamefont {A.~M.}\ \bibnamefont {Tsvelik}},\ }\bibfield  {title} {\bibinfo {title} {Multichannel {K}ondo impurity dynamics in a {M}ajorana device},\ }\href {https://doi.org/10.1103/PhysRevLett.113.076401} {\bibfield  {journal} {\bibinfo  {journal} {Phys. Rev. Lett.}\ }\textbf {\bibinfo {volume} {113}},\ \bibinfo {pages} {076401} (\bibinfo {year} {2014})}\BibitemShut {NoStop}%
\bibitem [{\citenamefont {Li}\ \emph {et~al.}(2023)\citenamefont {Li}, \citenamefont {K\"onig},\ and\ \citenamefont {V\"ayrynen}}]{Li2023}%
  \BibitemOpen
  \bibfield  {author} {\bibinfo {author} {\bibfnamefont {G.}~\bibnamefont {Li}}, \bibinfo {author} {\bibfnamefont {E.~J.}\ \bibnamefont {K\"onig}},\ and\ \bibinfo {author} {\bibfnamefont {J.~I.}\ \bibnamefont {V\"ayrynen}},\ }\bibfield  {title} {\bibinfo {title} {Topological symplectic {K}ondo effect},\ }\href {https://doi.org/10.1103/PhysRevB.107.L201401} {\bibfield  {journal} {\bibinfo  {journal} {Phys. Rev. B}\ }\textbf {\bibinfo {volume} {107}},\ \bibinfo {pages} {L201401} (\bibinfo {year} {2023})}\BibitemShut {NoStop}%
\bibitem [{\citenamefont {K\"onig}\ and\ \citenamefont {Tsvelik}(2023)}]{Koenig2023}%
  \BibitemOpen
  \bibfield  {author} {\bibinfo {author} {\bibfnamefont {E.~J.}\ \bibnamefont {K\"onig}}\ and\ \bibinfo {author} {\bibfnamefont {A.~M.}\ \bibnamefont {Tsvelik}},\ }\bibfield  {title} {\bibinfo {title} {Exact solution of the topological symplectic {K}ondo problem},\ }\href {https://doi.org/https://doi.org/10.1016/j.aop.2023.169231} {\bibfield  {journal} {\bibinfo  {journal} {Annals of Physics}\ }\textbf {\bibinfo {volume} {456}},\ \bibinfo {pages} {169231} (\bibinfo {year} {2023})}\BibitemShut {NoStop}%
\bibitem [{\citenamefont {Fiete}\ \emph {et~al.}(2008)\citenamefont {Fiete}, \citenamefont {Bishara},\ and\ \citenamefont {Nayak}}]{Fiete2008}%
  \BibitemOpen
  \bibfield  {author} {\bibinfo {author} {\bibfnamefont {G.~A.}\ \bibnamefont {Fiete}}, \bibinfo {author} {\bibfnamefont {W.}~\bibnamefont {Bishara}},\ and\ \bibinfo {author} {\bibfnamefont {C.}~\bibnamefont {Nayak}},\ }\bibfield  {title} {\bibinfo {title} {Multichannel {K}ondo models in non-{A}belian quantum {H}all droplets},\ }\href {https://doi.org/10.1103/PhysRevLett.101.176801} {\bibfield  {journal} {\bibinfo  {journal} {Phys. Rev. Lett.}\ }\textbf {\bibinfo {volume} {101}},\ \bibinfo {pages} {176801} (\bibinfo {year} {2008})}\BibitemShut {NoStop}%
\bibitem [{\citenamefont {K\"onig}\ \emph {et~al.}(2020)\citenamefont {K\"onig}, \citenamefont {Coleman},\ and\ \citenamefont {Tsvelik}}]{Koenig2020}%
  \BibitemOpen
  \bibfield  {author} {\bibinfo {author} {\bibfnamefont {E.~J.}\ \bibnamefont {K\"onig}}, \bibinfo {author} {\bibfnamefont {P.}~\bibnamefont {Coleman}},\ and\ \bibinfo {author} {\bibfnamefont {A.~M.}\ \bibnamefont {Tsvelik}},\ }\bibfield  {title} {\bibinfo {title} {Spin magnetometry as a probe of stripe superconductivity in twisted bilayer graphene},\ }\href {https://doi.org/10.1103/PhysRevB.102.104514} {\bibfield  {journal} {\bibinfo  {journal} {Phys. Rev. B}\ }\textbf {\bibinfo {volume} {102}},\ \bibinfo {pages} {104514} (\bibinfo {year} {2020})}\BibitemShut {NoStop}%
\bibitem [{\citenamefont {Matveev}(1991)}]{matveev1991}%
  \BibitemOpen
  \bibfield  {author} {\bibinfo {author} {\bibfnamefont {K.~A.}\ \bibnamefont {Matveev}},\ }\bibfield  {title} {\bibinfo {title} {Quantum fluctuations of the charge of a metal particle under the {C}oulomb blockade conditions},\ }\href {http://www.jetp.ras.ru/cgi-bin/e/index/e/72/5/p892?a=list} {\bibfield  {journal} {\bibinfo  {journal} {Soviet physics, JETP}\ }\textbf {\bibinfo {volume} {72}},\ \bibinfo {pages} {892} (\bibinfo {year} {1991})},\ \bibinfo {note} {{R}ussian original: {Zh. Eksp. Teor. Fiz}, {\bf 99}:5, 1598 (1991)}\BibitemShut {NoStop}%
\bibitem [{\citenamefont {Matveev}(1995)}]{matveev1995}%
  \BibitemOpen
  \bibfield  {author} {\bibinfo {author} {\bibfnamefont {K.~A.}\ \bibnamefont {Matveev}},\ }\bibfield  {title} {\bibinfo {title} {Coulomb blockade at almost perfect transmission},\ }\href {https://doi.org/10.1103/PhysRevB.51.1743} {\bibfield  {journal} {\bibinfo  {journal} {Phys. Rev. B}\ }\textbf {\bibinfo {volume} {51}},\ \bibinfo {pages} {1743} (\bibinfo {year} {1995})}\BibitemShut {NoStop}%
\bibitem [{\citenamefont {Furusaki}\ and\ \citenamefont {Matveev}(1995)}]{Furusaki1995}%
  \BibitemOpen
  \bibfield  {author} {\bibinfo {author} {\bibfnamefont {A.}~\bibnamefont {Furusaki}}\ and\ \bibinfo {author} {\bibfnamefont {K.~A.}\ \bibnamefont {Matveev}},\ }\bibfield  {title} {\bibinfo {title} {Theory of strong inelastic cotunneling},\ }\href {https://doi.org/10.1103/PhysRevB.52.16676} {\bibfield  {journal} {\bibinfo  {journal} {Phys. Rev. B}\ }\textbf {\bibinfo {volume} {52}},\ \bibinfo {pages} {16676} (\bibinfo {year} {1995})}\BibitemShut {NoStop}%
\bibitem [{\citenamefont {Iftikhar}\ \emph {et~al.}(2015)\citenamefont {Iftikhar}, \citenamefont {Jezouin}, \citenamefont {Anthore}, \citenamefont {Gennser}, \citenamefont {Parmentier}, \citenamefont {Cavanna},\ and\ \citenamefont {Pierre}}]{Iftikhar2015}%
  \BibitemOpen
  \bibfield  {author} {\bibinfo {author} {\bibfnamefont {Z.}~\bibnamefont {Iftikhar}}, \bibinfo {author} {\bibfnamefont {S.}~\bibnamefont {Jezouin}}, \bibinfo {author} {\bibfnamefont {A.}~\bibnamefont {Anthore}}, \bibinfo {author} {\bibfnamefont {U.}~\bibnamefont {Gennser}}, \bibinfo {author} {\bibfnamefont {F.~D.}\ \bibnamefont {Parmentier}}, \bibinfo {author} {\bibfnamefont {A.}~\bibnamefont {Cavanna}},\ and\ \bibinfo {author} {\bibfnamefont {F.}~\bibnamefont {Pierre}},\ }\bibfield  {title} {\bibinfo {title} {Two-channel {K}ondo effect and renormalization flow with macroscopic quantum charge states},\ }\href {https://doi.org/10.1038/nature15384} {\bibfield  {journal} {\bibinfo  {journal} {Nature}\ }\textbf {\bibinfo {volume} {526}},\ \bibinfo {pages} {233} (\bibinfo {year} {2015})}\BibitemShut {NoStop}%
\bibitem [{\citenamefont {Iftikhar}\ \emph {et~al.}(2018)\citenamefont {Iftikhar}, \citenamefont {Anthore}, \citenamefont {Mitchell}, \citenamefont {Parmentier}, \citenamefont {Gennser}, \citenamefont {Ouerghi}, \citenamefont {Cavanna}, \citenamefont {Mora}, \citenamefont {Simon},\ and\ \citenamefont {Pierre}}]{Iftikhar2018}%
  \BibitemOpen
  \bibfield  {author} {\bibinfo {author} {\bibfnamefont {Z.}~\bibnamefont {Iftikhar}}, \bibinfo {author} {\bibfnamefont {A.}~\bibnamefont {Anthore}}, \bibinfo {author} {\bibfnamefont {A.~K.}\ \bibnamefont {Mitchell}}, \bibinfo {author} {\bibfnamefont {F.~D.}\ \bibnamefont {Parmentier}}, \bibinfo {author} {\bibfnamefont {U.}~\bibnamefont {Gennser}}, \bibinfo {author} {\bibfnamefont {A.}~\bibnamefont {Ouerghi}}, \bibinfo {author} {\bibfnamefont {A.}~\bibnamefont {Cavanna}}, \bibinfo {author} {\bibfnamefont {C.}~\bibnamefont {Mora}}, \bibinfo {author} {\bibfnamefont {P.}~\bibnamefont {Simon}},\ and\ \bibinfo {author} {\bibfnamefont {F.}~\bibnamefont {Pierre}},\ }\bibfield  {title} {\bibinfo {title} {Tunable quantum criticality and super-ballistic transport in a {``charge''} {K}ondo circuit},\ }\href {https://doi.org/10.1126/science.aan5592} {\bibfield  {journal} {\bibinfo  {journal} {Science}\ }\textbf {\bibinfo {volume} {360}},\ \bibinfo {pages} {1315} (\bibinfo {year} {2018})}\BibitemShut {NoStop}%
\bibitem [{\citenamefont {Karki}\ \emph {et~al.}(2022)\citenamefont {Karki}, \citenamefont {Boulat},\ and\ \citenamefont {Mora}}]{Karki2022}%
  \BibitemOpen
  \bibfield  {author} {\bibinfo {author} {\bibfnamefont {D.~B.}\ \bibnamefont {Karki}}, \bibinfo {author} {\bibfnamefont {E.}~\bibnamefont {Boulat}},\ and\ \bibinfo {author} {\bibfnamefont {C.}~\bibnamefont {Mora}},\ }\bibfield  {title} {\bibinfo {title} {Double-charge quantum island in the quasiballistic regime},\ }\href {https://doi.org/10.1103/PhysRevB.105.245418} {\bibfield  {journal} {\bibinfo  {journal} {Phys. Rev. B}\ }\textbf {\bibinfo {volume} {105}},\ \bibinfo {pages} {245418} (\bibinfo {year} {2022})}\BibitemShut {NoStop}%
\bibitem [{\citenamefont {Pouse}\ \emph {et~al.}(2021)\citenamefont {Pouse}, \citenamefont {Peeters}, \citenamefont {Hsueh}, \citenamefont {Gennser}, \citenamefont {Cavanna}, \citenamefont {Kastner}, \citenamefont {Mitchell},\ and\ \citenamefont {Goldhaber-Gordon}}]{Pouse2023}%
  \BibitemOpen
  \bibfield  {author} {\bibinfo {author} {\bibfnamefont {W.}~\bibnamefont {Pouse}}, \bibinfo {author} {\bibfnamefont {L.}~\bibnamefont {Peeters}}, \bibinfo {author} {\bibfnamefont {C.~L.}\ \bibnamefont {Hsueh}}, \bibinfo {author} {\bibfnamefont {U.}~\bibnamefont {Gennser}}, \bibinfo {author} {\bibfnamefont {A.}~\bibnamefont {Cavanna}}, \bibinfo {author} {\bibfnamefont {M.~A.}\ \bibnamefont {Kastner}}, \bibinfo {author} {\bibfnamefont {A.~K.}\ \bibnamefont {Mitchell}},\ and\ \bibinfo {author} {\bibfnamefont {D.}~\bibnamefont {Goldhaber-Gordon}},\ }\bibfield  {title} {\bibinfo {title} {Quantum simulation of an exotic quantum critical point in a two-site charge {K}ondo circuit},\ }\href {https://doi.org/10.48550/ARXIV.2108.12691} {\bibfield  {journal} {\bibinfo  {journal} {arXiv}\ }\textbf {\bibinfo {volume} {2108}},\ \bibinfo {pages} {12691} (\bibinfo {year} {2021})},\ \bibinfo {note} {to appear in {N}ature {P}hysics}\BibitemShut {NoStop}%
\end{thebibliography}
\end{document}